\documentclass[useAMS,usenatbib,usegraphicx]{mn2e}
\usepackage{times}

\def\dd{{\rm{d}}}
\def\mh{\hbox{[M/H]}}
\def\dex{\,{\rm dex}}
\def\mag{\,{\rm mag}}
\def\cM{{\cal M}}
\def\selfn{P(S | \bar{\mathbf{y}}, \mathbf{x}, \bsigma_y)}
\def\thirdwidth{63mm}

\title[Stellar distances from spectroscopic observations]{Stellar distances from spectroscopic observations: a new technique}
\author[B. Burnett and J. Binney]{Benedict Burnett\thanks{E-mail: burnett@thphys.ox.ac.uk} and James Binney\\
Rudolf Peierls Centre for Theoretical Physics, Keble Road, Oxford, OX1 3NP, UK}
\begin{document}

\date{22 April 2010}

\pagerange{\pageref{firstpage}--\pageref{lastpage}} \pubyear{2010}

\maketitle

\label{firstpage}

\begin{abstract}
A Bayesian approach to the determination of stellar distances from photometric
and spectroscopic data is presented and tested both on pseudodata, 
designed to mimic data for stars observed by the RAVE survey, and on the real stars from the Geneva-Copenhagen survey. It is argued
that this method is optimal in the sense that it brings to bear all available
information and that its results are limited only by observational errors and
the underlying physics of stars. The method simultaneously returns the
metallicities, ages and masses of programme stars. Remarkably, the
uncertainty in the output metallicity is typically 44 per cent smaller than
the uncertainty in the input metallicity.
\end{abstract}


\section{Introduction}

Present attempts to reconcile models of the Galaxy with observations are
hampered by a lack of full 6d phase space data for a large number of stars.
While the majority of recent stellar surveys -- for example the
Geneva-Copenhagen survey (\citealt{Nordstrom}) and the Radial Velocity Experiment
\citep[RAVE;][]{Steinmetz:RAVE} -- provide sky positions, proper motions and often
radial velocities, the persistent difficulty is the determination of
distances, without which proper-motion measurements are of limited value.
Since the fitting of models to observational constraints is arguably the
method with the most potential for understanding the true nature of the
Galaxy, the lack of these distance data is a key obstacle to our
understanding of the Milky Way, and thus to our understanding of galaxies in
general.

Several methods of distance determination are known. Perhaps most famously
and successfully, the Hipparcos satellite measured trigonometric parallaxes
for $\sim\! 10^5$ stars down to a \textit{V}-band magnitude of around 12
(\citealt{Hipparcos}). Trigonometric parallax is conceptually the most
fundamental distance measurement technique for stars; however the Hipparcos
measurements were accurate only out to a distance of around 200\,pc. Within
the next decade, the Gaia mission (\citealt{Gaia}) should return parallaxes
for around $10^9$ stars, dramatically increasing the size of the available
data set for investigation; however in the meantime trigonometric parallaxes
to a vast number of otherwise well-observed stars are lacking. Furthermore, even after
the Gaia mission is complete, other estimation methods will remain vital for
more distant stars.

Two other trigonometric methods of distance measurement are known: so-called
`Galactic parallax' (\citealt{Galpar,Galpar2}) and `geometrodynamical'
techniques (\citealt{LyndenBelldistance}).  However, both these methods
require strong constraints on the orbits of target stars, so they are
restricted to a small subset of all stars.

The next most reliable distance determination technique is the use of
so-called `standard candles' -- stars with particular properties such as
RR~Lyrae and Cepheid variables, whose luminosities are a reasonably
sharply-defined function of observables such as period and colour. While the
accuracies attainable for such stars may be high, standard candles form an
extremely small subset of any  stellar population. Hence although they can
be used to give a broad-brush picture of Galactic structure (see
\citealt{Gautschy}), the technique cannot be applied to the majority of stars
of interest.

The other major technique for distance estimation is the determination of
`photometric distances'. This involves deducing the luminosity of a star from
its colours and perhaps metallicity, allowing its distance to be inferred
from its apparent magnitude. \cite{Juric_cut} have applied this technique to
millions of stars measured by the Sloan Digital Sky Survey (hereafter SDSS).
The technique works well for SDSS stars because they are apparently faint and
therefore overwhelmingly likely to lie on the main sequence,
and the luminosity of such a star is a well-defined function of colour and
metallicity. However for samples at brighter apparent magnitudes it is not
safe to assume that stars lie on the main sequence, and a more sophisticated technique is
required. 

An example of such a sample is the RAVE survey, which is in many ways
complementary to the SDSS sample. It covers a nominal $I$-band magnitude
range of $9 < I < 12$ and is expected to provide spectrophotometric data on
up to a million stars by 2011 (\citealt{RaveDR1_cut}). On account of its
magnitude range, the RAVE sample contains a non-negligible number of
giants. Such a heterogeneous sample therefore requires a more sophisticated
approach to parameter determination than a direct colour-to-absolute
magnitude mapping. Furthermore it would be much more satisfactory to obtain a
methodology that can give a definitive distance estimation (and corresponding
individual uncertainty) for specific stars rather than only for sizeable
statistical samples.

Recently \cite{Breddels} estimated photometric distances to $\sim\! 25\,000$
stars in the RAVE survey. These distances were obtained by repeatedly seeking
the stellar model that provides the best fit to first the data and then
$5\,000$ pseudo-data obtained by scattering the observables of the first
stellar model by the observational errors. The logical justification of this
procedure is unclear. In this paper we argue that the principles of Bayesian
inference lead unambiguously to a different procedure that has much in common
with the procedures for determining stellar ages introduced by
\cite{PontEyer} and \cite{Jorgensen}. In fact, our procedure yields not only
a distance to each star but also estimates of its mass, age and metallicity.
We argue that our procedure is optimal in the sense that it exploits the
available data in their entirety -- including information about the survey in
question. From these facts it constructs a pdf for each star in model space,
so estimated distances, masses, ages and metallicities are accompanied by error
estimates.  The method is applicable to any survey that provides more than
one non-degenerate observable for each star and is limited only by
degeneracies in the underlying stellar physics and any inherent inaccuracies
in the stellar models.

The paper is structured as follows: Section~\ref{sec:theory} covers the
theoretical basis of the method. Section~\ref{sec:pseudodata} explores the
application of the method to a fake data set.
Section~\ref{sec:GCS} then looks at the results of its application to the Geneva-Copenhagen sample.
Section~\ref{sec:previouswork} details the relationship of this paper to
previous. Section~\ref{sec:conclusions} presents a 
discussion of the results and highlights the potential for future
applicability.

\section[]{Theory} \label{sec:theory}

Let $\mathbf{y}$ represent an $n$-tuple of a star's
observable quantities (e.g.\ effective temperature $T_{\rm{eff}}$, surface
gravity $\log g$, observed metallicity $\mh_{\rm{obs}}$, colours, apparent
magnitudes, sky position) and let $\mathbf{x}$ represent its six `intrinsic'
variables (true initial metallicity $\mh$, age $\tau$, initial mass $\cM$,
distance $s$, and true sky position $(l,b)$) -- i.e.\
 \begin{equation}
  \mathbf{x} = \left( \mh, \tau, \cM, s, l, b \right).
\end{equation}
We will assume that a stellar model can be used to provide a direct mapping
from $\mathbf{x}$-space to $\mathbf{y}$-space. For the purpose of concision,
let $G(\mathbf{w},\bsigma)$ represent the multivariate Gaussian function 
 \begin{equation}
  G(\mathbf{w},\bsigma) \equiv \prod_{i=1}^n \left( \sigma_i \sqrt{2\pi}\right)^{-1} \exp \left( - w_i^2 / 2 \sigma_i^2 \right)
\end{equation}
for an $n$-tuple $\mathbf{w}$.

In order to estimate a value and an uncertainty for each stellar parameter,
the natural approach to take is probabilistic. We know a set of facts: the
actual measured values of a star's observables ($\bar{\mathbf{y}}$), the quoted errors thereon ($\bsigma_y$) and,
subtly, the fact that the star is in the given sample (let this fact be
denoted by $S$). Hence the logical distribution to consider is the pdf of a
star's parameters $x_i$ given these three facts. If we can find this
distribution, then we have all the parametric information that can logically
be inferred from the known facts -- most importantly, we can give an
expectation value for each parameter and an estimated error on this value.

Explicitly, for each component $x_i$ of $\mathbf{x}$ we seek the three moments
$\mathcal{I}_{ik}$ defined by
 \begin{equation} \label{eq:firstmoments}
  \mathcal{I}_{ik} = \int \! x_i^k \,\, p(\mathbf{x} | \bar{\mathbf{y}}, \bsigma_y , S ) \, \dd^6 \mathbf{x} ,
\end{equation}
where $k \in \{0,1,2\}$. (The $k=0$ moment should properly be unity, but it is useful if the pdf is left unnormalized.) From these moments we can then infer the expectation and variance of each stellar parameter.

In order to express the pdf in terms of known distributions, it is useful to
consider the full pdf
 \begin{eqnarray} \label{eq:fullpdf}
  p(\mathbf{x}, \bar{\mathbf{y}}, \bsigma_y , S) &\!\! =& \!\! p(\mathbf{x} | \bar{\mathbf{y}}, \bsigma_y , S) \, p(\bar{\mathbf{y}}, \bsigma_y , S) \\
  &\!\! =& \!\! P(S | \bar{\mathbf{y}}, \mathbf{x}, \bsigma_y) \, p(\bar{\mathbf{y}} | \mathbf{x}, \bsigma_y ) \, p(\bsigma_y | \mathbf{x}) \, p(\mathbf{x})
  .\nonumber
\end{eqnarray}
 Hence we can expand the pdf in a form analogous to Bayes' theorem to give
 \begin{equation}
  p(\mathbf{x} | \bar{\mathbf{y}}, \bsigma_y , S) = \frac{ P(S | \bar{\mathbf{y}}, \mathbf{x}, \bsigma_y) \, p(\bar{\mathbf{y}} | \mathbf{x}, \bsigma_y ) \, p(\bsigma_y | \mathbf{x}) \, p(\mathbf{x}) }{ p(\bar{\mathbf{y}}, \bsigma_y , S) } ,
\end{equation}
which can be simplified to
\begin{equation} \label{eq:firsteq}
  p(\mathbf{x} | \bar{\mathbf{y}}, \bsigma_y , S)  \propto P(S | \bar{\mathbf{y}}, \mathbf{x}, \bsigma_y) \, p(\bar{\mathbf{y}} | \mathbf{x}, \bsigma_y ) \, p(\bsigma_y | \mathbf{x}) \, p(\mathbf{x}) ,
\end{equation}
wherein factors independent of $\mathbf{x}$ have been neglected as
irrelevant to the problem we aim to solve, since we will require only ratios of
different $\mathcal{I}_{ik}$'s.

Each of the factors in equation~(\ref{eq:firsteq}) can now be analysed
independently:
\begin{enumerate}
\item $\selfn$, which we term the \textit{selection function},
expresses the probability of a star being in the sample given its inherent
values and the measurement errors. This factor therefore reflects our
selection criteria, whether they be, for instance, a magnitude cut or any cut
on errors used to `clean' a sample.

\item For many types of observation the \textit{likelihood} $p(\bar{\mathbf{y}} |
\mathbf{x}, \bsigma_y )$ can be approximated by a Gaussian
$G(\bar{\mathbf{y}} - \mathbf{y}(\mathbf{x}), \bsigma_y)$. More generally it
takes a functional form determined purely by the measuring instrument, in which
the different components of $\bar{\mathbf{y}}$ may or may not be independent.

\item $p(\bsigma_y | \mathbf{x})$ is the probability of the quoted errors in
$\mathbf{y}$ given the  object's underlying characteristics $\mathbf{x}$.

\item $p(\mathbf{x})$ is our \textit{prior}. This will describe as many
stellar populations as we wish to take into account, in terms of an initial
mass function (IMF) and a spatial distribution. It describes the relative
abundance of stars of various types, without regard to their observability.

\end{enumerate}

The uncertainties on the observed values of $(l,b)$ are assumed to be
sufficiently small for us to regard the corresponding pdfs as delta functions
and drop the integrals over $l$ and $b$, leaving a 4-dimensional integral in
$\mathbf{x}$. However, when this is done, any locational factors in the prior
must gain a multiplicative factor of $s^2$ in order to take account of the
conical shape of the volume surveyed.

In this work we assume a uniform distribution for $p(\bsigma_y |
\mathbf{x})$; although in reality this will not be the case (for instance,
more distant stars will tend to have higher errors), the dependence of
$\bsigma_y$ on $\mathbf{x}$ will generally be sufficiently weak not to affect
our results greatly. In this case, equation~(\ref{eq:firsteq}) simplifies to
 \begin{equation}\label{eq:final}
  p(\mathbf{x} | \bar{\mathbf{y}}, \bsigma_y , S) \propto P(S | \bar{\mathbf{y}}, \mathbf{x}, \bsigma_y) \, p(\bar{\mathbf{y}} | \mathbf{x}, \bsigma_y ) \, p(\mathbf{x}) .
\end{equation}
This can be substituted into the
simplified form of equation~(\ref{eq:firstmoments}),
\begin{equation}
  \mathcal{I}_{ik} = \int \! x_i^k \, p(\mathbf{x} | \bar{\mathbf{y}}, \bsigma_y , S ) \, \dd^4 \mathbf{x} ,
\end{equation}
 giving a set of three integrals that together will give us an estimated
parameter value for each star ($\left< x_i \right> = \mathcal{I}_{i1}
/\mathcal{I}_{i0}$) and an uncertainty thereon ($\sigma_i =
\sqrt{\left(\mathcal{I}_{i2} /\mathcal{I}_{i0}\right) - \left< x_i
\right>^2}$), taking into account all of the known information. Thus in a
single pass we can infer all these values for each star's metallicity, age,
mass and distance.

\subsection{Selection function}

The nature of the selection function $P(S | \bar\mathbf{y}, \mathbf{x},
\bsigma_y)$ bears comment, as it is by no means trivial. We must distinguish
carefully what is meant by the symbol $\bar\mathbf{y}$ in the formalism
above: it refers to the actual observed parameters of the star, but only
those we actually use in the analysis. There may be other parameters that
have been observed and enter into the selection function. If this is the
case, they can be expressed only as probabilities dependent on $\mathbf{x}$,
not on $\bar\mathbf{y}$. Therefore it becomes important to split the
selection function into two parts, one dependent on $\bar\mathbf{y}$ and the
other on $\mathbf{x}$:
\begin{equation}
P(S |
\bar\mathbf{y}, \mathbf{x}, \bsigma_y) = \psi(\bar\mathbf{y}, \bsigma_y)
\,\phi(\mathbf{x}).
\end{equation}
 In this product $\psi$ will generally be the dominant factor for two main
reasons:

\begin{itemize}
\item[(i)]
If the sample is based on an input catalogue, it will frequently be a
selection on observables that are not then reobserved -- for example, the
Geneva-Copenhagen survey (\citealt{Nordstrom}) was selected on the basis of the
Str\"omgren photometry of \cite{Olsen1,Olsen2,Olsen3,Olsen4}, which is simply
transcribed to the Geneva-Copenhagen catalogue.

\item[(ii)] The fundamental limitations on stellar photometry also tend to
make $\psi$ dominant: saturation leads to a bright-end cut being imposed on
the survey; at the faint end a cutoff is imposed to eliminate
objects too faint for the detector's source extraction algorithm to
distinguish true signals from noise. Both cuts are generally imposed on the
basis of the observed magnitude and therefore are functions of the observed
properties of a star \textit{after scattering by observational errors}. That
is, they are functions of $\bar\mathbf{y}$ rather than $\mathbf{x}$.

\end{itemize}
However, there can be contributions of the nature of $\phi(\mathbf{x})$, and
these are of much greater importance to our analysis. For example, we will
later be interested in comparison of the results provided by our method on a
real stellar sample with those of the Hipparcos satellite. In such cases, it
is common (e.g.\ \citealt{Breddels}) to consider only those stars for which
Hipparcos provides reasonably well-constrained distances, by restricting the
sample to stars with fractional parallax errors below a certain threshold. It
is important to recognize that this induces a bias in the sample: by
considering only those stars with, say, $\sigma_\varpi / \overline\varpi <
20\%$ (where $\overline\varpi$ represents the observed parallax and
$\sigma_\varpi$ the error thereon), one will preferentially cut away stars
with low parallaxes and hence large distances, leaving a sample biased
towards small distances.

This effect can, however, be at least mitigated by a consideration of the
r\^{o}le of $\phi(\mathbf{x})$. For any value of $\mathbf{x}$, and hence a
`true' model distance $s$, it is possible to define a pdf for a star's
observed parallax $\overline\varpi$ from a knowledge of $\sigma_\varpi$,
assuming Gaussian errors: indeed
 \begin{equation}
  p\left( \overline\varpi | s \right) = G \left( \overline\varpi -1/s , \sigma_\varpi \right) .
\end{equation}
 Therefore one can incorporate this bias into one's analysis in the form
 \begin{equation}
\phi(\mathbf{x}) \propto \int_{5 \sigma_\varpi}^\infty \!\! G \left( \overline\varpi - 1/s , \sigma_\varpi \right) \, \dd \overline\varpi ,
\end{equation}
 providing an element of balance in the analysis that maximum-likelihood
techniques neglect.

If one can, as will generally be the case at least approximately, decompose the
selection function into the form $\psi(\bar\mathbf{y}, \bsigma_y)
\,\phi(\mathbf{x})$, then since we are interested only in terms dependent on
$\mathbf{x}$, we can ignore any occurrence of $\psi(\bar\mathbf{y},
\bsigma_y)$. Hence our final formula for the moments of a star's pdf
collapses to
 \begin{equation}\label{eq:integrate}
  \mathcal{I}_{ik} = \int \! x_i^k \, \phi(\mathbf{x}) \,p(\bar{\mathbf{y}} | \mathbf{x}, \bsigma_y ) \, p(\mathbf{x}) \, \dd^4 \mathbf{x} ,
\end{equation}
which can be readily calculated.

\subsection{Implementation}

Since for many surveys the errors on apparent magnitude are very small, the
integration of equation~(\ref{eq:integrate}) need only cover a small range in
distance for each set of $(\mh,\tau,\cM)$ values. Consequently, after
experimenting with both fixed-grid and Markov Chain Monte Carlo techniques,
it was decided that the best compromise between speed and reliability was
provided by a simple fixed-grid Newton-Cotes rectangle integration method, where
one defines a grid of points in metallicity, the logarithm of age, and mass.
For each point in this three-dimensional space one then integrates over a
range in distance corresponding to an apparent magnitude spread of say ten
times the magnitude error either side of the observed value -- any distances
beyond this range will give a negligible contribution to the integral due to
the Gaussian factor in the likelihood term $p(\bar{\mathbf{y}} | \mathbf{x},
\bsigma_y )$. For the purposes of calculating the integral it was found to
be sufficient simply to multiply the integrand's value at each grid point by
an effective volume determined by the distance between the given grid point
and its nearest neighbours.

No two numbers completely characterize a general probability
distribution. Medians, means and modes can all be used to characterize the
centre of the distribution. The median is the most stable measure and
generally to be preferred. Unfortunately, it is in general hard to calculate.
The mode, which was favoured by \cite{Jorgensen}, is susceptible to noise
when the distribution is at all flat-topped. Therefore we favour the mean as
a robust and readily calculated measure of the pdf. Similarly, we favour the
variance as an estimate of the width of the pdf, rather than the more complex
confidence intervals used by \citeauthor{Jorgensen}.

Another advantage of using these two numbers to characterize the distribution
is that their calculation, via the integration technique outlined above,
avoids the thorny issue of interpolation on the isochrones by considering only
the values of observables at tabulated grid points. Interpolation using 
isochrones is notoriously difficult, and thus we avoid it as far as is possible.

To summarize, the method we propose runs as follows:
\begin{enumerate}
\item Define a grid of points in $(\mh,\log\tau,\cM)$ space. We found a suitable 
grid spacing to be given by the metallicity values given in Table~\ref{table:Zs}
(although a greater range would do no harm -- one should include the lowest 
metallicity that is available from the isochrones); a spacing in $\log\tau$ of 0.025; 
and the mass spacing provided automatically by the Padova isochrones, which ranges
from $10^{-6}$\,M$_\odot$ to 0.255\,M$_\odot$.
 \item At each grid point, define a further set of points in distance $s$,
covering the range 
 \begin{equation}\label{eq:distance}
  s \in [r(m-n\sigma_m), r(m+n\sigma_m)],
\end{equation}
 where $m$ is the apparent magnitude used for fitting to observations,
$\sigma_m$ is the observational error thereon, and $r(m)$ represents the
distance that produces an apparent magnitude of $m$ given the metallicity,
age and mass at the grid point in question. $n$ is some positive number
chosen by the user -- we found $n=5$ to be ample for high
precision.

 \item Assign to each point a 4-volume corresponding to a
hypercuboid with boundary planes halfway to each of the point's nearest
neighbours (at the edges of parameter space, take the length in the limiting
dimension to be that between the final two data points). 

 \item Perform the integral in equation~(\ref{eq:integrate}) by finding the value of
$\phi(\mathbf{x}) \, 
p(\bar{\mathbf{y}} | \mathbf{x}, \bsigma_y ) \, p(\mathbf{x})$ at each grid point,
multiplied by the grid point's associated 4-volume. We term this product of
probability density and volume the `weight' of each grid point. For each of
the first four dimensions of $\mathbf{x}$, one can then define an array that
keeps track of the sum of these weights multiplied by powers of each
parameter $x_i$, and also keep a running total of the sum of the weights
themselves, which we term $\mathcal{I}_0$. 

 \item Once one has run over the entire grid (or some subset thereof, since
one may be able to disregard metallicities too far from the observed value),
one can then calculate the moments $\mathcal{I}_{ik}$ by dividing each
running total by $\mathcal{I}_0$. From the $\mathcal{I}_{ik}$ one then has
directly an expectation and uncertainty for each dimension of $\mathbf{x}$.

\item It is of course imperative to check that one has sampled $\mathbf{x}$-space sufficiently
finely to achieve sufficient precision in the work. For these purposes one should
redo the analysis of a sample of stars using a finer sampling in each dimension
and check that there is minimal alteration in the outputs for the new sampling. If
time is not a concern, one would ideally do this for every star, refining the grid 
iteratively to a point where there is minimal change in the outputs. This may 
require interpolation if the isochrones cannot be provided with a fine enough sampling.
\end{enumerate}

\subsection{Terminology\label{sec:terminology}}

In what follows, we shall refer to the technique described above as the
`Bayesian method', in order to distinguish it from what we shall call the
`maximum-likelihood method'. The distinction here essentially concerns the
nature of the prior and selection function. Maximum-likelihood techniques for
fitting a model to data consist of finding the model (from some specified
range) that maximizes the probability of the data in question -- i.e.\ in our
case one would seek to maximize the likelihood $p(\bar{\mathbf{y}} |
\mathbf{x}, \bsigma_y )$. This is mathematically equivalent to setting both
the selection function and the prior in the above formalism to uniform
distributions, but conceptually it has an important difference from our
Bayesian approach, which seeks to find the model with the maximum probability
of being correct given the data. Maximum-likelihood techniques are popular
due to their simplicity -- they involve no consideration of the nature of the
prior and they generally do not
take into account the selection function. Furthermore, from a conceptual
viewpoint, the Bayesian technique is a more justified approach to the problem
of model selection than maximum likelihood: since the data are given, it is
logical to seek the model with the maximum probability of being correct given
the data. For the case of determining stellar ages, \cite{Jorgensen} give a
nice example of the advantage of a Bayesian approach over maximum likelihood,
while \cite{PontEyer} provide an extensive discussion of the relative merits
of a Bayesian technique.

One more point bears comment with regard to maximum likelihood (and na{\"i}ve
Bayesian approaches) when the prior is set to be uniform in order to
represent an unprejudiced starting assumption. For a uniform continuous pdf,
the prior can only be defined to be uniform in a specific coordinate system:
a transformation of those coordinates will in general not leave it so. Thus,
for example, while a uniform prior in age may seem a reasonable starting
point, it is difficult to justify such an assumption as opposed to a prior
uniform in e.g.\ the logarithm of age. Thus the assumption of a uniform prior
in some space is nonetheless an assumption, and cannot be considered safer
than making an explicit choice of prior after considering all the
circumstances of the particular case. In general one hopes for the likelihood
function to be sufficiently strongly peaked at some value to render the exact
form of the prior unimportant for the posterior distribution; however in such
complex, degenerate cases as those addressed here involving stellar
evolution, such hopes are not always well-founded. For this reason a
prior that is based on what we do know of stars in the Galaxy is an
important factor in any calculation.

\section{Test case} \label{sec:pseudodata}

\subsection{Sample}

In order to test the consistency of the method, a fake data set was
generated to mimic the sample observed by RAVE. For these purposes, the
vector of observables was taken to be 
 \begin{equation}
  \mathbf{y} = \left( \log T_{\rm{eff}}, \log g, \mh_{\rm{obs}}, J-K, J, I, l, b \right),
\end{equation}
where $I$, $J$ and $K$ denote apparent magnitudes. (Here, and throughout this
paper, logarithms are taken to base 10.) Stars were generated by a Markov
Chain following the Metropolis-Hastings algorithm (\citealt{Saha}) in
$\mathbf{x}$-space, with observational errors added in
$\mathbf{y}$-space at every step. Since the aim was to reproduce the joint pdf described by equation~(\ref{eq:fullpdf}), at every proposed point in $\mathbf{x}$-space, a $\bar\mathbf{y}$ value was generated by Gaussian scattering by a vector of observational errors, and the probability $\selfn$ was calculated; the acceptance was then based on the new value of $\selfn \, p(\mathbf{x})$.

The Markov chain was generated 
using a Gaussian proposal density, and ensuring an acceptance rate of order
30\% by tuning the spread of the proposal density during a burn-in period of
10\,000 steps. The chain itself consisted of $2.5 \times 10^6 $ steps, of which forty-nine out
of every fifty were then discarded (in order to minimize the chance of repeated 
$\mathbf{x}$-points) to provide the output 50\,000 stars. Convergence
was verified by performing an identical run with the selection function switched
off, and ensuring that the marginalized distributions in $\mathbf{x}$ corresponded to
those input in the prior.

The Gaussian observational errors were added using an error 8-tuple
 \begin{equation}
  \bsigma_y = \left( 0.0434, \gamma(\bar\mathbf{y}_1), 1.07 \,\bar\mathbf{y}_1 - 3.71, 0.045, 0.023, 0.04, 0, 0 \right),
\end{equation}
where
\begin{equation}
  \gamma(x) = \cases{0.5 & if $x<\log (8\,000)$,\cr
    0.25 + 0.436 \left(x - \log (8\,000) \right) & otherwise;}
\end{equation}
 designed to be representative of the scale of observational errors in RAVE.
Errors on logarithmic quantities are measured in dex.  (Although the true
errors on $\log T_{\rm{eff}}$ and other derived observables may not actually
be Gaussian, it was considered to be a sufficient approximation for this
proof-of-concept.  A different error distribution could be employed very
easily.)  The pseudodata were therefore generated according to a distribution
function described by
 \begin{equation} \label{eq:fake}
  f(\mathbf{x}, \bar\mathbf{y}) = p(\mathbf{x}, \bar\mathbf{y}, \bsigma_y |
  S) \propto \selfn \, p(\bar{\mathbf{y}} | \mathbf{x}, \bsigma_y ) \, p(\mathbf{x}) ,
\end{equation}
 where the factors on the right are as follows:

For the
prior we took a three-component Milky Way model of the form 
\begin{equation}\label{eq:priorofx}
  p(\mathbf{x}) = p(\cM) \sum_{i=1}^3 p_i(\mh) \, p_i(\tau) \, p_i(\mathbf{r}),
\end{equation}
where $i=1,2,3$ correspond to a thin disc, thick disc and stellar halo,
respectively. We assumed an identical Kroupa-type IMF for all three
components and distinguish them as follows:

\paragraph*{Thin disc ($i=1$):}
\begin{eqnarray} \label{eq:thindisc}
  p_1(\mh) &=& G(\mh, 0.2), \nonumber \\
  p_1(\tau)  &\propto& \exp(0.119 \,\tau/\mbox{Gyr}) \quad \mbox{for $\tau \leqslant 10$\,Gyr,}  \\
  p_1(\mathbf{r}) &\propto& \exp\left(-\frac{R}{R_d^{\rm{thin}}} - \frac{|z|}{z_d^{\rm{thin}}}  \right);  \nonumber
\end{eqnarray}

\paragraph*{Thick disc ($i=2$):}
\begin{eqnarray}
  p_2(\mh) &=& G(\mh+0.6, 0.5), \nonumber \\
  p_2(\tau)  &\propto& \mbox{uniform in range $8 \leqslant \tau \leqslant 12$\,Gyr,} \\
  p_2(\mathbf{r}) &\propto& \exp\left(-\frac{R}{R_d^{\rm{thick}}} - \frac{|z|}{z_d^{\rm{thick}}}  \right); \nonumber
\end{eqnarray}

\paragraph*{Halo ($i=3$):}
\begin{eqnarray}
  p_3(\mh) &=& G(\mh+1.6, 0.5), \nonumber \\
  p_3(\tau)  &\propto& \mbox{uniform in range $10 \leqslant \tau \leqslant 13.7$\,Gyr,} \\
  p_3(\mathbf{r}) &\propto& r^{-3.39}; \nonumber
\end{eqnarray}
 where $R$ signifies Galactocentric cylindrical radius, $z$ cylindrical
height and  $r$ spherical radius. The parameter values were taken as
in Table~\ref{table:params}; the values
are taken from the analysis of SDSS data in \cite{Juric_cut}. The metallicity
and age distributions for the thin disc come from \cite{Haywood} and
\cite{Aumer}, while the radial density of the halo comes from the `inner
halo' detected in \cite{Carollo}. The metallicity and age distributions of
the thick disc and halo are influenced by \cite{Reddy} and \cite{Carollo}.

The normalizations were then adjusted so that at the solar position, taken as
$R_0=$~8.33\,kpc (\citealt{Gillessen}), $z_0=$~15\,pc, we have number
density ratios $n_2 /n_1 = 0.15$ (\citealt{Carollo}), $n_3 /n_1 = 0.005$
(\citealt{Juric_cut}).

\begin{table}
  \begin{center}
    \caption{Values of disc parameters used.\label{table:params}}
    \begin{tabular}{cr}
      \hline
      Parameter & Value (pc) \\
      \hline
      $R_d^{\rm{thin}}$  & 2\,600 \\[3pt]
      $z_d^{\rm{thin}}$  & 300 \\[3pt]
      $R_d^{\rm{thick}}$ & 3\,600 \\[3pt]
      $z_d^{\rm{thick}}$ & 900 \\
      \hline
    \end{tabular}
  \end{center}
\end{table}

The IMF chosen follows the form originally proposed by \cite{Kroupa}, with a
minor modification following \cite{Aumer}, being
 \begin{equation}
  p(\cM) \propto \cases{\cM^{-1.3}&if $\cM<0.5\,$M$_\odot$,\cr
    0.536 \, \cM^{-2.2}&if $0.5\,$M$_\odot \leqslant \cM<1\, $M$_\odot$,\cr
    0.536 \, \cM^{-2.519}&otherwise.}
\end{equation}

We determined $\mathbf{y}$ as a function of
$\mathbf{x}$ from  the isochrones of the Padova group
(\citealt{Bertelli}), which provide tabulated values for the observables of
stars with metallicities ranging upwards from around $\mh \approx -0.92$,
ages in the range $\tau \in [0.01, 19]$\,Gyr and masses in the range $\cM \in
[0.15, 20]$\,M$_\odot$. We used isochrones for 16 metallicities as shown in
Table~\ref{table:Zs}, selecting the helium mass fraction $Y$ as a function of
metal mass fraction $Z$ according to the relation used in \cite{Aumer}, i.e.\
$Y \approx 0.225 + 2.1 Z$ and assuming solar values of $(Y_\odot , Z_\odot) =
(0.260,0.017)$. The metallicity values were selected by eye to ensure that there 
was not a great change in the stellar observables between adjacent isochrone sets.

\begin{table}
  \begin{center}
    \caption{Metallicities of isochrones used, taking $(Z_\odot , Y_\odot) = (0.017,0.260)$.\label{table:Zs}}
    \begin{tabular}{rrr}
      \hline
      $Z$ & $Y$ & $\mh$ \\
      \hline
      0.0022 & 0.230 & $-0.914$ \\
      0.003 & 0.231 & $-0.778$ \\
      0.004 & 0.233 & $-0.652$ \\
      0.006 & 0.238 & $-0.472$ \\
      0.008 & 0.242 & $-0.343$ \\
      0.010 & 0.246 & $-0.243$ \\
      0.012 & 0.250 & $-0.160$ \\
      0.014 & 0.254 & $-0.090$ \\
      0.017 & 0.260 & 0.000 \\
      0.020 & 0.267 & 0.077 \\
      0.026 & 0.280 & 0.202 \\
      0.036 & 0.301 & 0.363 \\
      0.040 & 0.309 & 0.417 \\
      0.045 & 0.320 & 0.479 \\
      0.050 & 0.330 & 0.535 \\
      0.070 & 0.372 & 0.727 \\
      \hline
    \end{tabular}
  \end{center}
\end{table}

The selection function $\selfn$ was chosen to  describe RAVE's
selection criteria. RAVE observes stars with nominal DENIS $I$-band
magnitudes in the range $9 < I_{\rm DENIS} < 12$; however \cite{RaveDR2_cut}
explain that the upper limit actually extends up to one magnitude fainter,
and that there is evidence of saturation around $I_{\rm DENIS} < 10$. For
these reasons it was decided to take the full range of $I$-band magnitudes
observable by RAVE to be $4<\bar{I}<13$, and to disallow stars falling outside this
range. Although the brighter limit may seem overly permissive, its actual value
has little importance due to the other major factor in $\selfn$: a completeness term. Although RAVE is
theoretically capable of observing all stars within its magnitude limits, it
is not a complete survey and thus stars at certain magnitudes have a higher
probability of being included in the catalogue than others. For the purposes
of this test, it was decided to use an approximation of fig.~4 of
\cite{RaveDR1_cut}, of the form
 \begin{equation}
  \selfn \propto 2.9 \,G(\bar{I}-9.8,0.76) + G(\bar{I}-11.7,0.51) .
\end{equation}
 While this neglects variations in completeness with sky position, it seems a
reasonable approximation for our pseudodata. The functional form results in
stars with particularly bright apparent magnitudes being given very low
weight irrespective of our chosen value for RAVE's low-$\bar{I}$ cutoff.

The other factor included was a cutoff at Galactic latitudes of $|b|
\leqslant 25^\circ$, since RAVE avoids regions close to the Galactic plane.
This has the obvious effect of biasing the sample slightly towards thick-disc
and halo stars. (It also prevents any density divergence near the Galactic 
centre due to the halo density profile.)

\subsection{Results}

\begin{figure}
  \includegraphics[width=84mm]{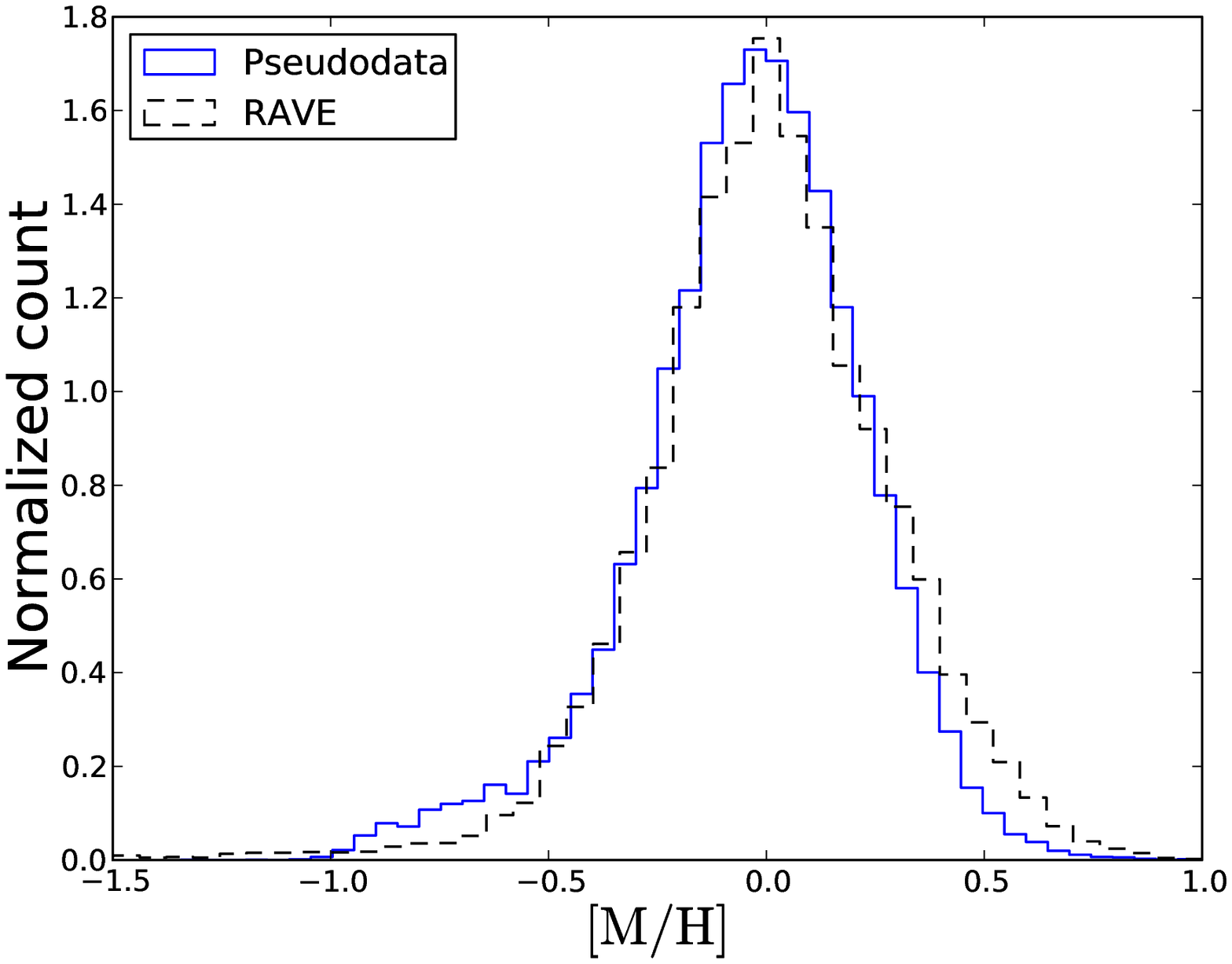}
  \includegraphics[width=84mm]{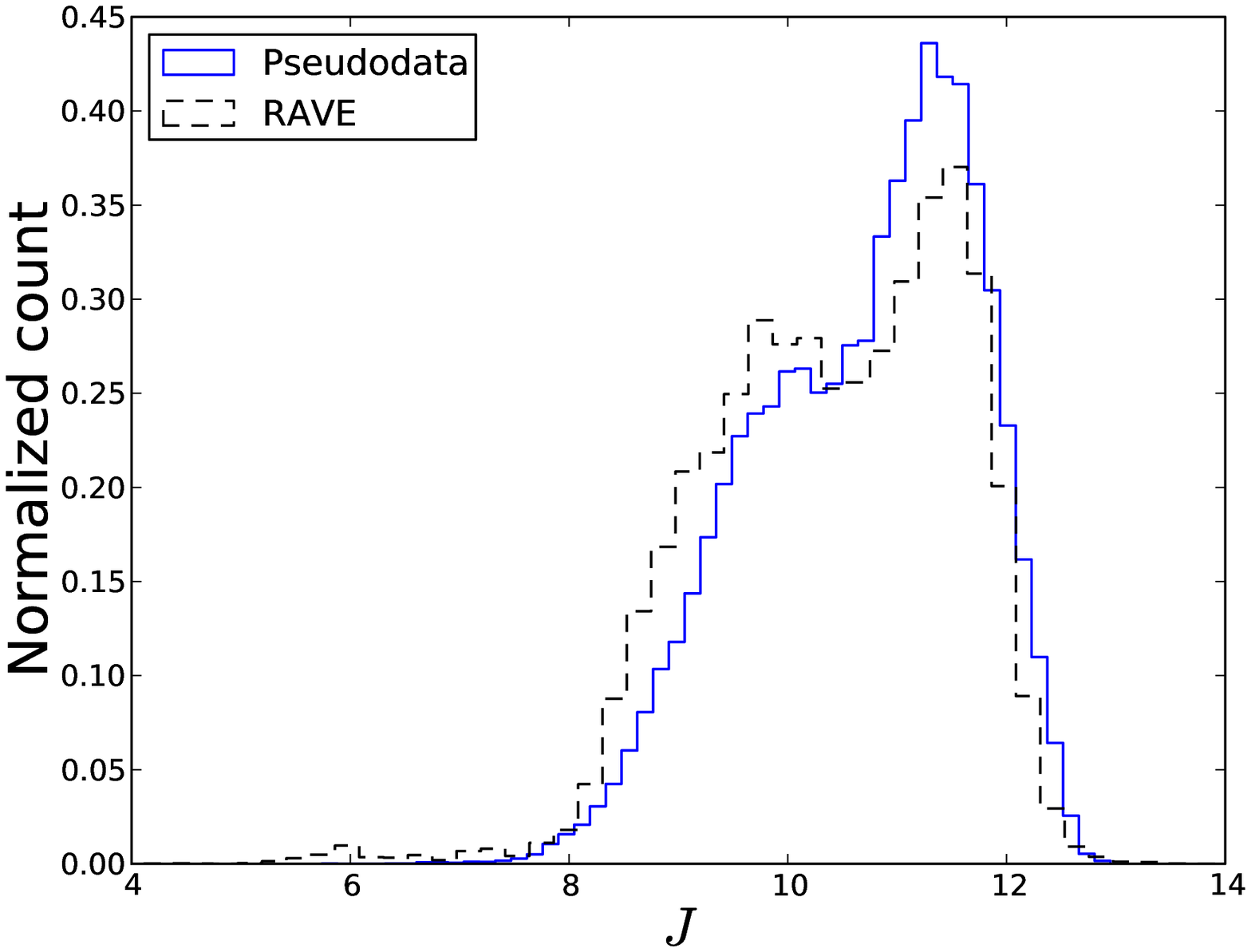}
  \caption{Distribution in metallicity (upper plot) and $J$-magnitude (lower
plot) for the pseudodata (full lines) and  stars in the RAVE catalogue 
(dashed lines).}
  \label{fig:ZJ}
\end{figure}

50\,000 stars were generated according to the above model. Interpolation in
the isochrones was necessary for these purposes; since the chosen sampling of
the isochrones was reasonably dense it was decided to use linear
interpolation rather than a more complex and arbitrary method. The density of
the sampling points should ensure that any errors introduced by this
interpolation technique are minimal, and certainly sufficiently small for the
present testing purpose.  The resultant distributions in metallicity and in
$J$ magnitude are displayed in Fig.~\ref{fig:ZJ}, along with that of the
second data release of the RAVE survey (\citealt{RaveDR2_cut}), showing that
the sample is a reasonable mimic. This sample was then analysed using the
technique expounded in Section~\ref{sec:theory}. The $\mathbf{y}$-space
fitting (the likelihood term $p(\bar{\mathbf{y}} | \mathbf{x}, \bsigma_y )$)
was performed in the first five components of $\mathbf{y}$. It was decided
against using the $I$-band for analysis due to the saturation concerns
described above,
and the consequent fear that the $I$-magnitudes of model stars will not correspond
to the $I_{\rm DENIS}$ system. The same reliability fear militates against
considering $I(\mathbf{x})$ in the factor $\phi(\mathbf{x})$. 
The factor $p(\mathbf{x})$ was initially given the same form
for the analysis as for the sample generation, while the factor
$\phi(\mathbf{x})$ was taken to be flat, since the selection function was
entirely a function of $\bar\mathbf{y}$.

In order to perform the integration of equation~(\ref{eq:integrate}), we
followed the prescription of Section~\ref{sec:theory}. After trying various
subdivisions of $\mathbf{x}$-space, it was determined that the optimal
subdivision, balancing speed against accuracy, was obtained by taking the
gridpoints of the integration at the metallicity values found in
Table~\ref{table:Zs}, increments of 0.025 in $\log (\tau /{\rm{Gyr}})$, and
the non-uniform mass spacing provided  by the isochrones.
Integration in these three dimensions was performed over the entire isochrone
range. The distance integration was performed using $n=5$ in
equation~(\ref{eq:distance}) for the apparent $J$-magnitude, with 5 distance
points used for each point in $(\mh,\log\tau,\cM)$. Using more points than
this was found to give negligible improvement in results, since the actual
spread for each $(\mh,\log\tau,\cM)$ value is extremely small due to the tiny
values of $\sigma_J$ in the 2MASS survey ($\sim\! 0.023$\,mag,
\citealt{2MASS_cut}), and correspondingly in the RAVE sample.

\begin{figure}
  \includegraphics[width=84mm]{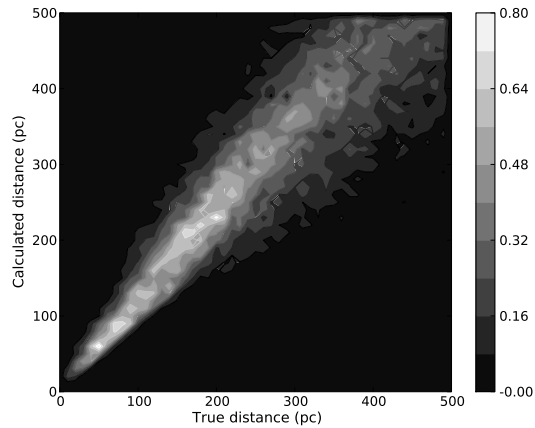}
  \caption{Density plot of the relationship of calculated to true distance for stars out to 500\,pc. Contours measure density in points~${\rm pc}^{-2}$.}
  \label{fig:scatter}
\end{figure}

\begin{figure}
  \centering
  \includegraphics[width=\thirdwidth]{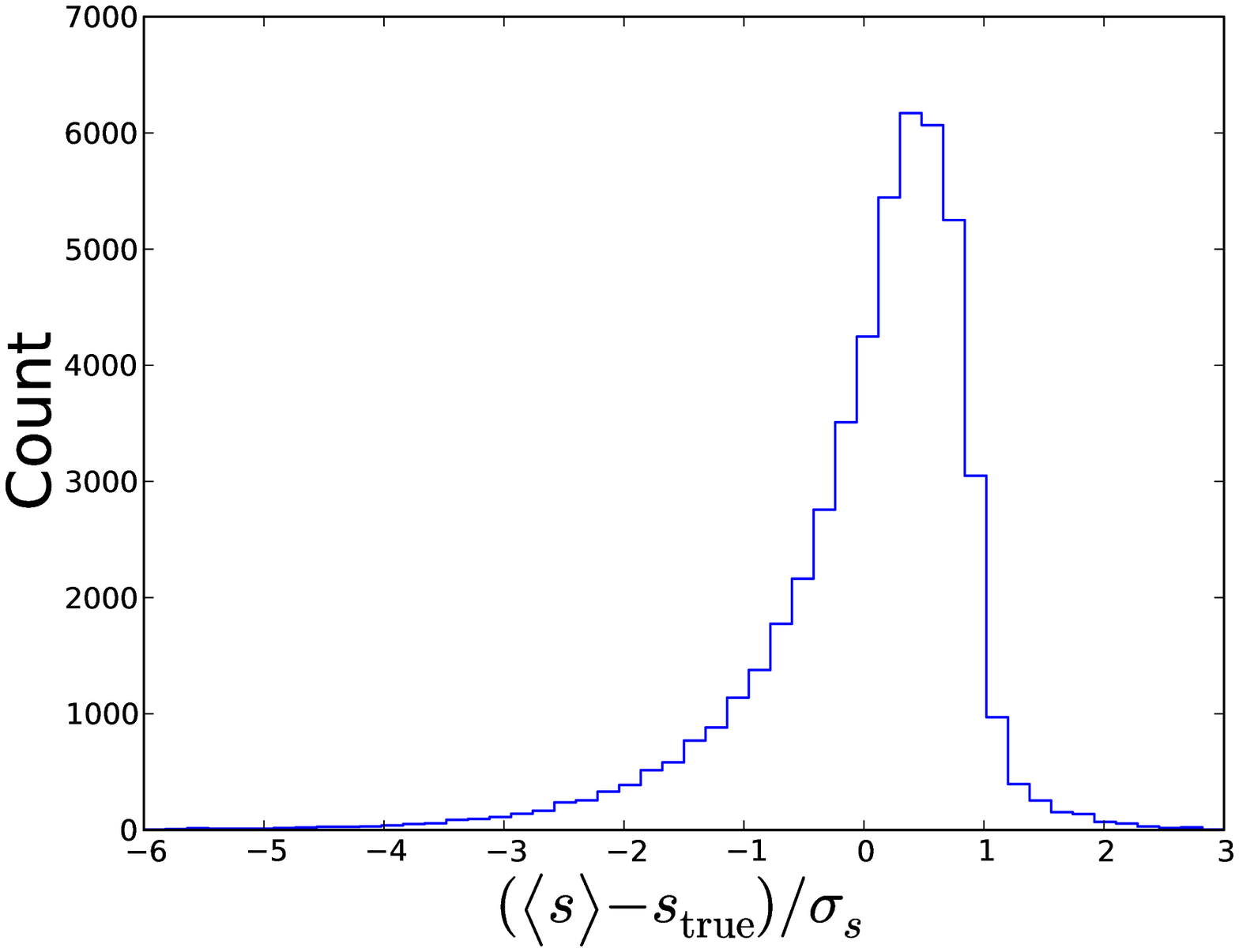}
  \includegraphics[width=\thirdwidth]{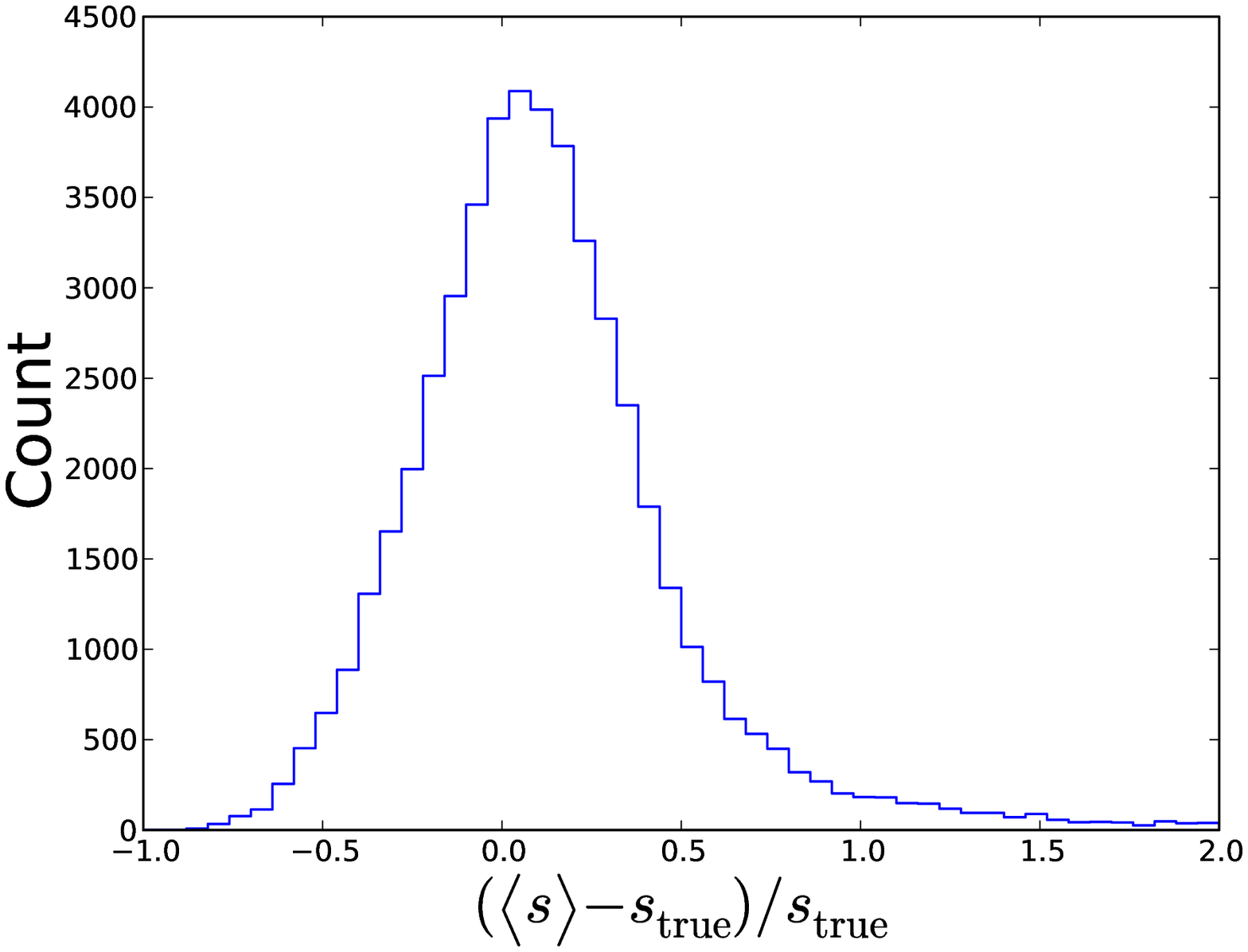}
  \includegraphics[width=\thirdwidth]{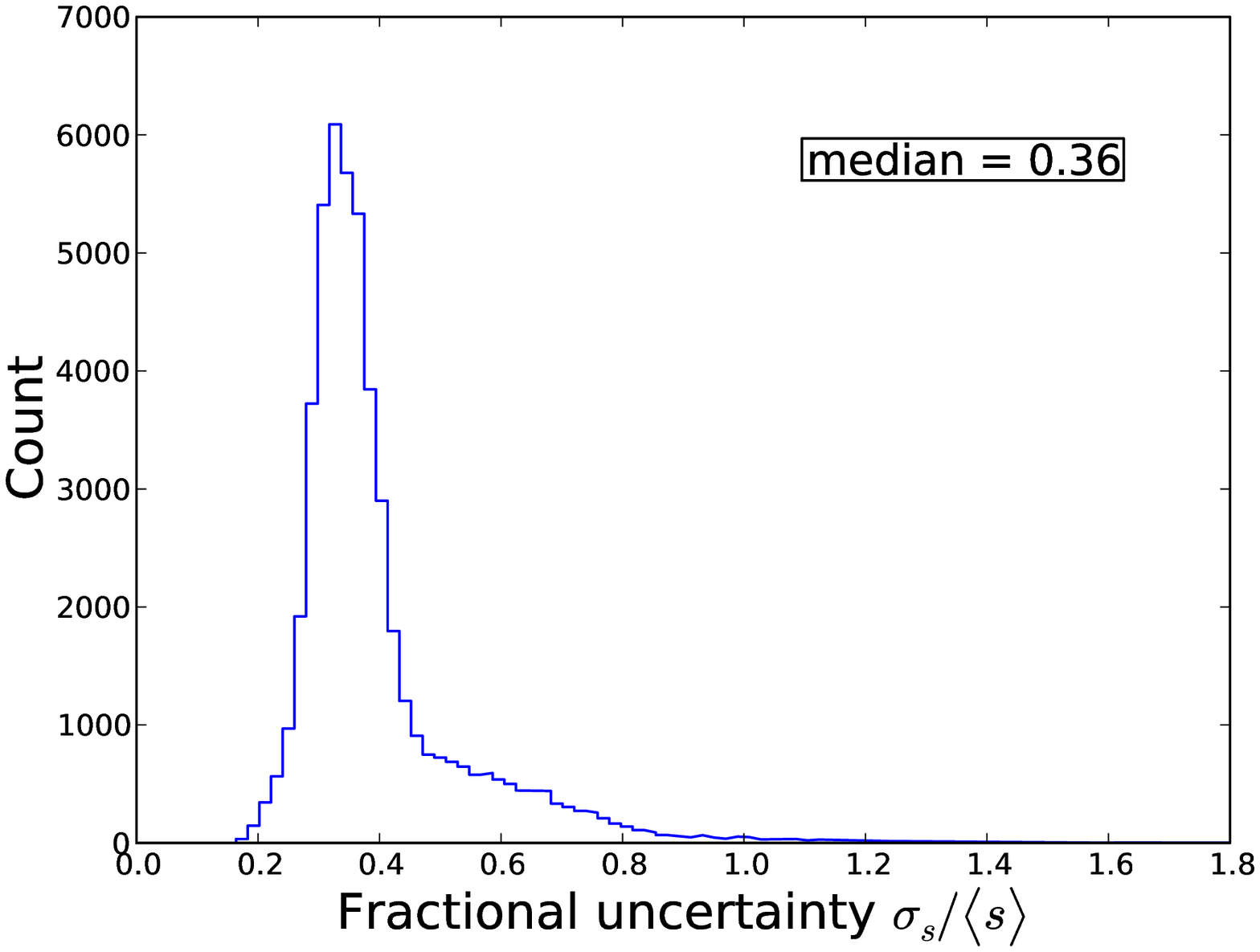}
  \caption{Results of analysis using the Bayesian method. Top panel:
    distribution of normalized residuals. While the distribution is skewed, it is
    almost entirely unbiased. Middle panel: distribution of residuals as a 
    fraction of true distance.
    Bottom panel: distribution of fractional uncertainties.}
  \label{fig:studentized}  
\end{figure}

\begin{figure}
  \centering
  \includegraphics[width=\thirdwidth]{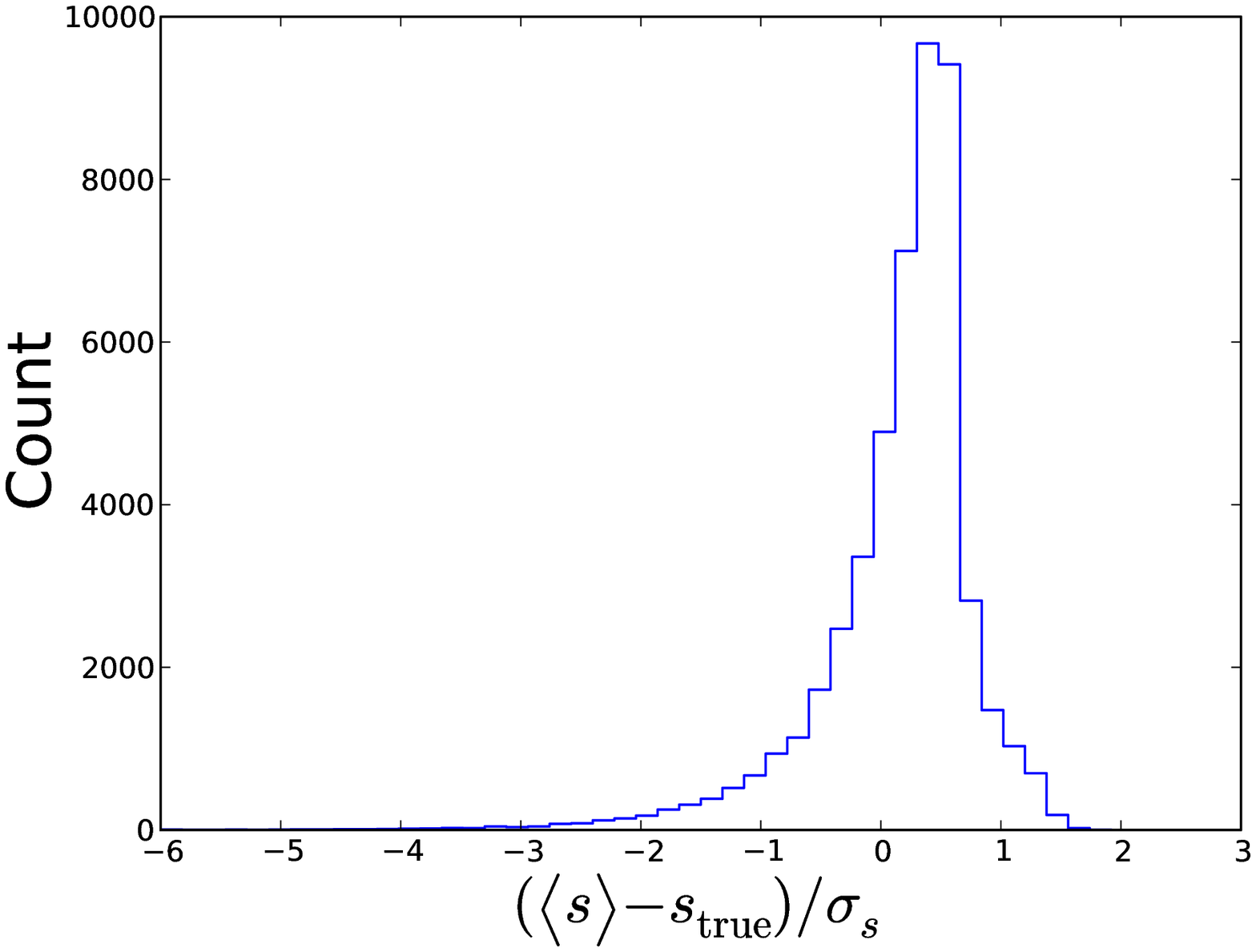}
  \includegraphics[width=\thirdwidth]{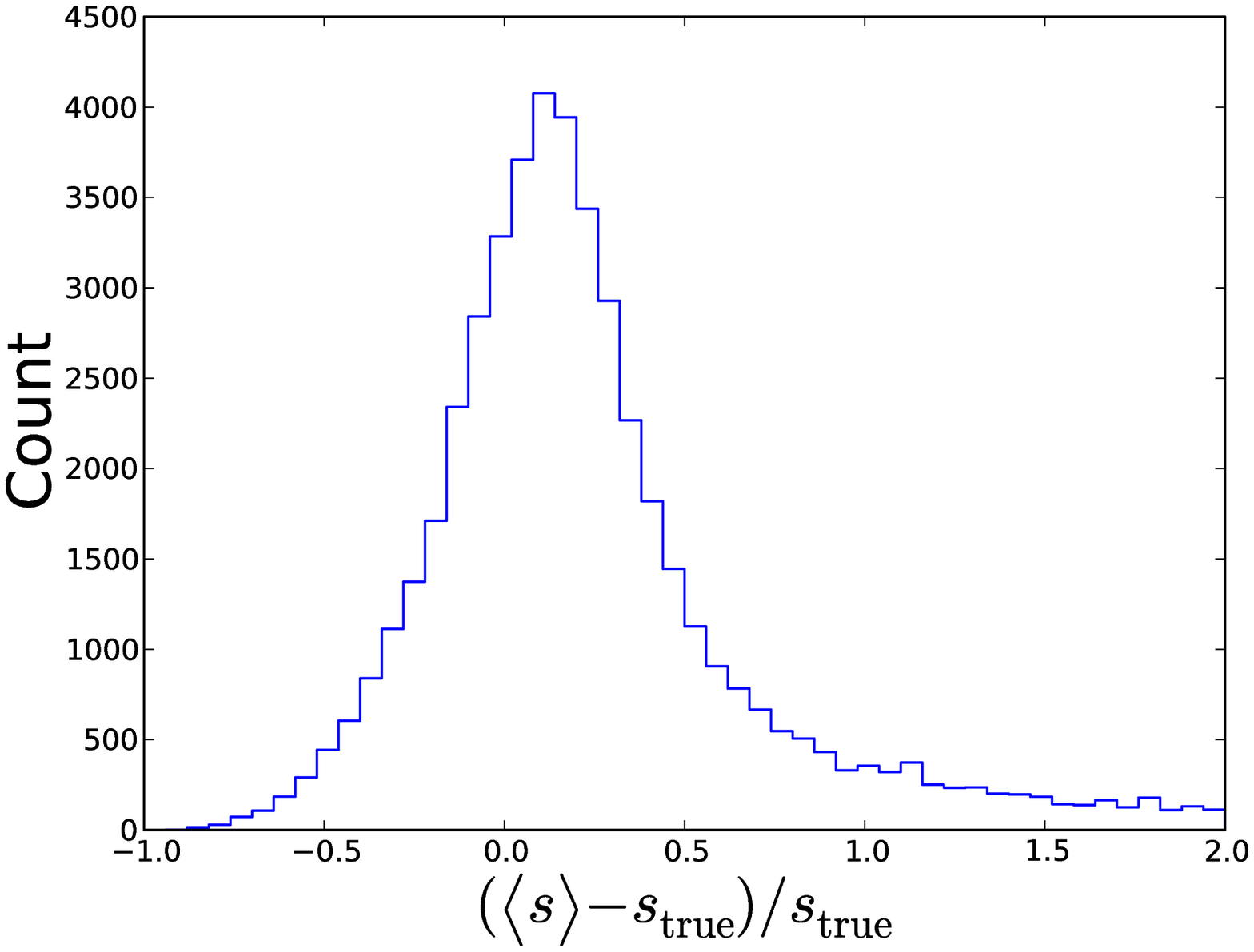}
  \includegraphics[width=\thirdwidth]{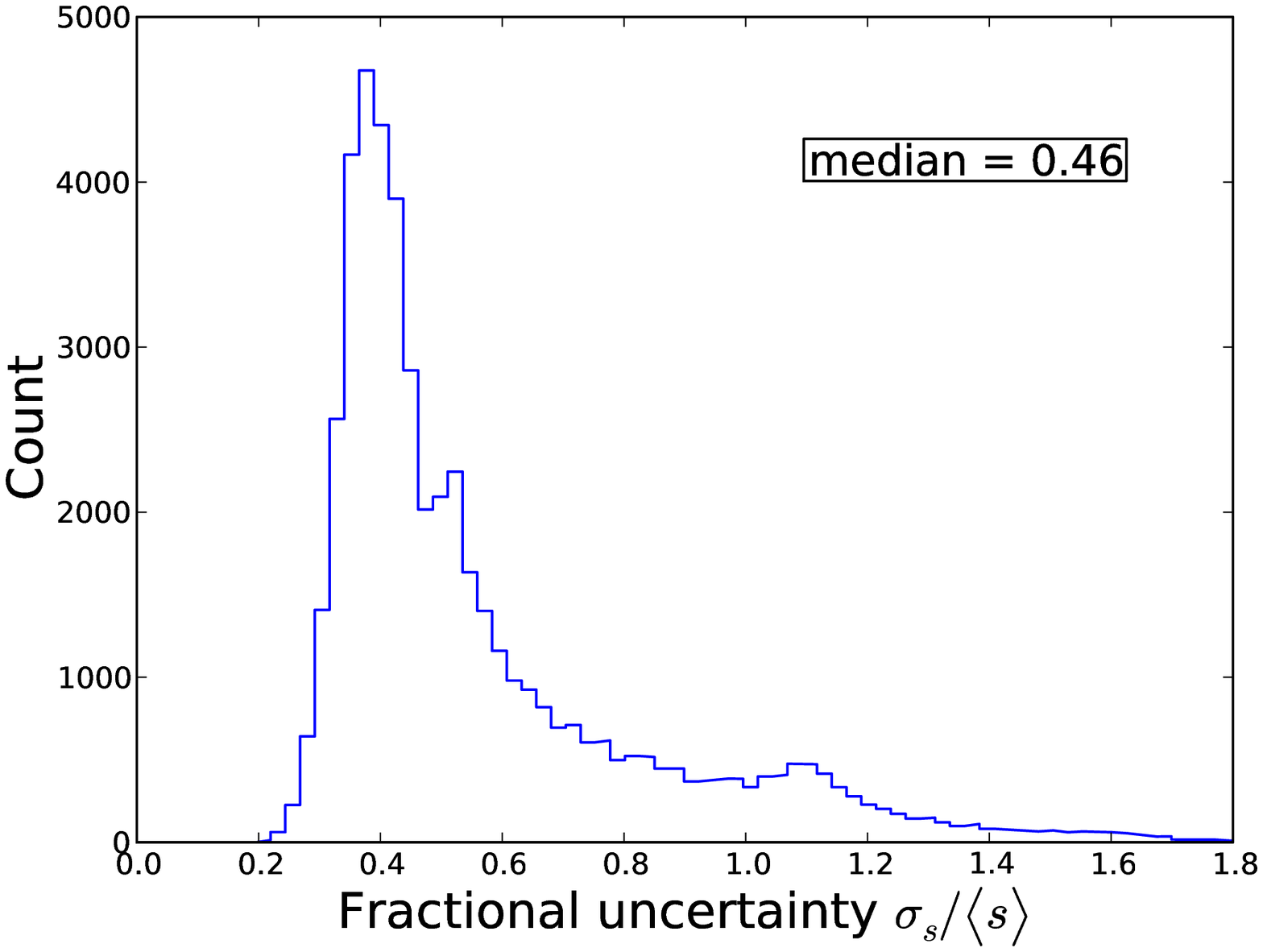}
  \caption{As Fig.~\ref{fig:studentized}, but for a maximum-likelihood
    method.}
  \label{fig:maxlik}
\end{figure}

The results of the analysis are displayed in Figs.~\ref{fig:scatter} and
\ref{fig:studentized}. Fig.~\ref{fig:studentized} displays a histogram of the
difference between the calculated value of each star's distance and its true
value, divided by the distance uncertainty $\sigma_s$ returned by the method.
Although it can be clearly seen that the distribution is skewed (indeed we
have no a priori reason to expect it not to be), it is \textit{not} biased:
the distribution displayed has a mean value of $0.009$ and a dispersion of
$0.93$, giving a reasonable sign that individual error estimates are
trustworthy. For comparison, a fit using a simple maximum-likelihood method
(i.e.\ dropping the factor $p(\mathbf{x})$ from the analysis) is displayed in
Fig.~\ref{fig:maxlik}.  The mean of the distribution shown in the top panel
of Fig.~\ref{fig:maxlik} is $0.15$, with a dispersion of $0.68$; this bias
confirms the criticism of flat priors in Section~\ref{sec:terminology}. It is
noteworthy that this bias persists despite the fact that the output
uncertainties are significantly larger in the maximum-likelihood case, as
demonstrated by the bottom panel of Fig.~\ref{fig:maxlik} (indeed, the small
dispersion of the top panel of this figure implies that these uncertainties
are systematically overestimated). Furthermore, the positive wing displayed
in the middle panel of Figs.~\ref{fig:studentized} and \ref{fig:maxlik} is
notably more pronounced in the maximum-likelihood case. Hence it can be seen
that the method outstrips standard photometric distance determination
techniques.

In the case of an analysis of real stellar survey data, the prior will not be
known to perfect precision. Consequently, we have also performed the analysis
with two different incorrect priors, each consisting of a single stellar
population:

\paragraph*{Approximate prior 1:}
\begin{eqnarray}
  p(\mh) &=& G(\mh+0.12, 0.2), \nonumber \\
  p(\tau)  &\propto& \exp(0.119 \,\tau/\mbox{Gyr}) \quad \mbox{for $\tau \leqslant 10$\,Gyr,}  \\
  p(\cM)  &\propto& \cM^{-2.35}, \nonumber \\
  p(\mathbf{r}) &\propto& \exp\left(-\frac{R}{2\,000\,{\rm pc}} - \frac{|z|}{400\,{\rm pc}}  \right).  \nonumber
\end{eqnarray}

\paragraph*{Approximate prior 2:}
\begin{eqnarray}
  p(\mh) &=& G(\mh, 0.3), \nonumber \\
  p(\tau)  &\propto& \exp(0.13 \,\tau/\mbox{Gyr}) \quad \mbox{for $\tau \leqslant 10$\,Gyr,}  \\
  p(\cM)  &\propto& \cM^{-2.35}, \nonumber \\
  p(\mathbf{r}) &\propto& \exp\left(-\frac{R}{2\,000\,{\rm pc}} - \frac{|z|}{400\,{\rm pc}}  \right).  \nonumber
\end{eqnarray}

\begin{figure}
  \centering
  \includegraphics[width=\thirdwidth]{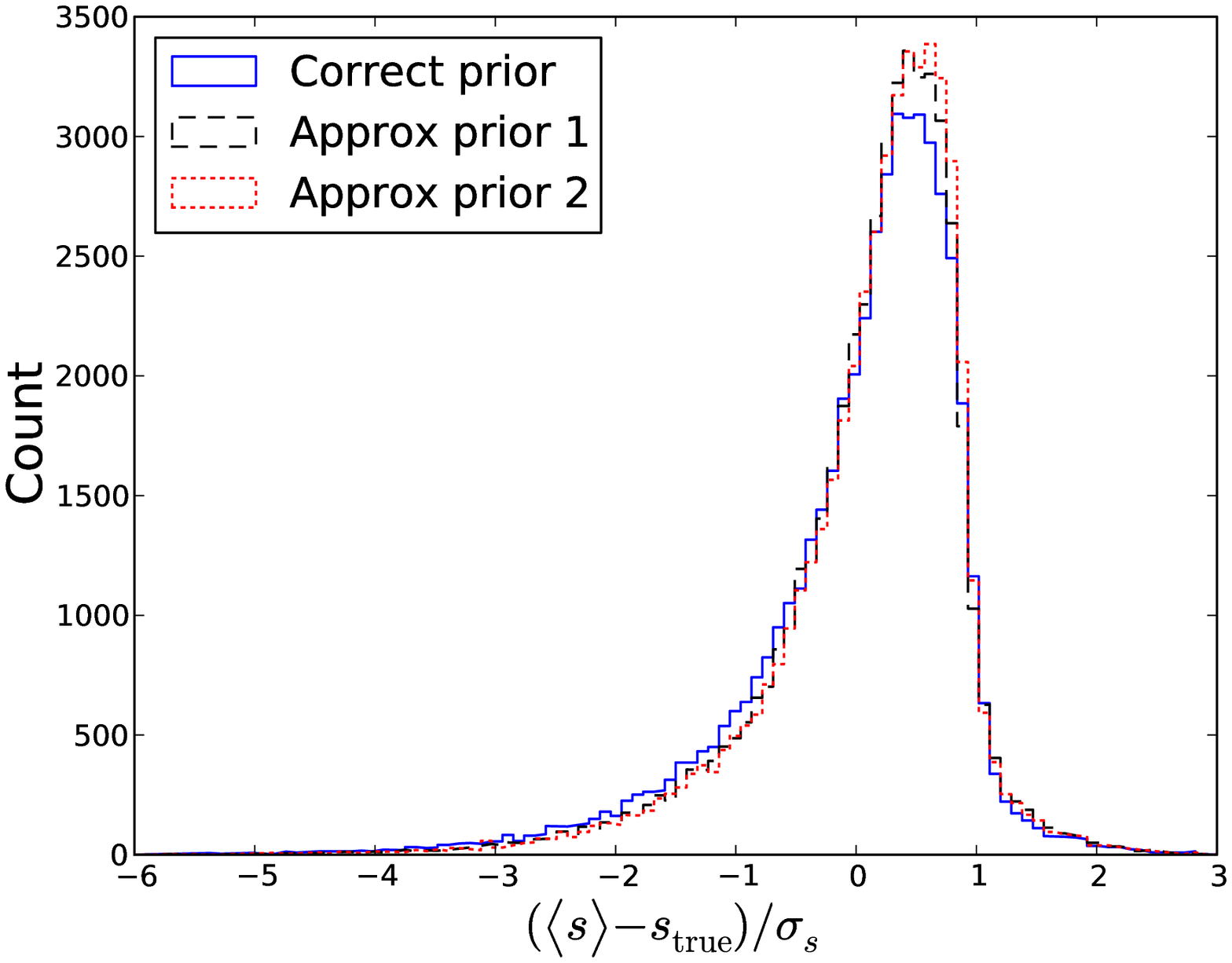}
  \includegraphics[width=\thirdwidth]{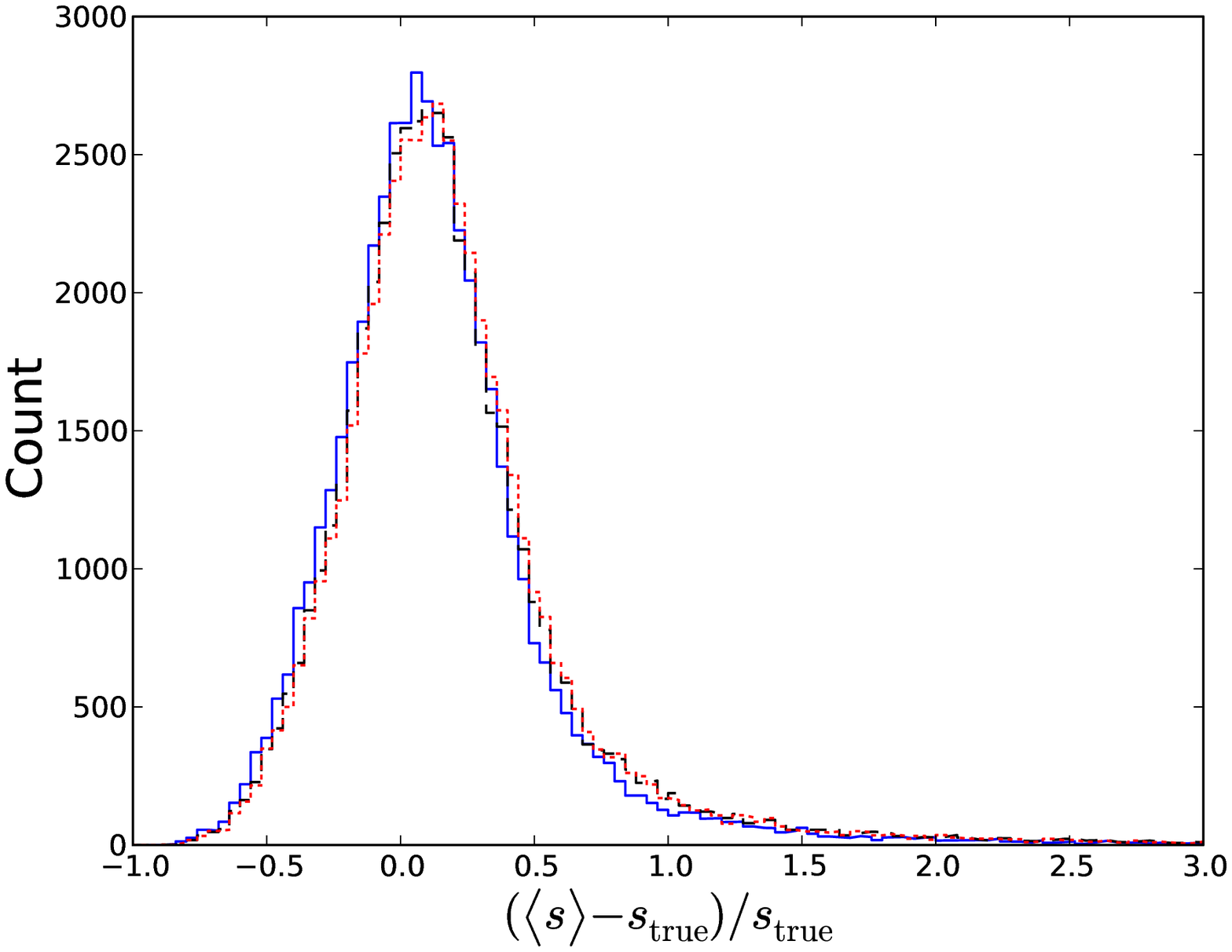}
  \includegraphics[width=\thirdwidth]{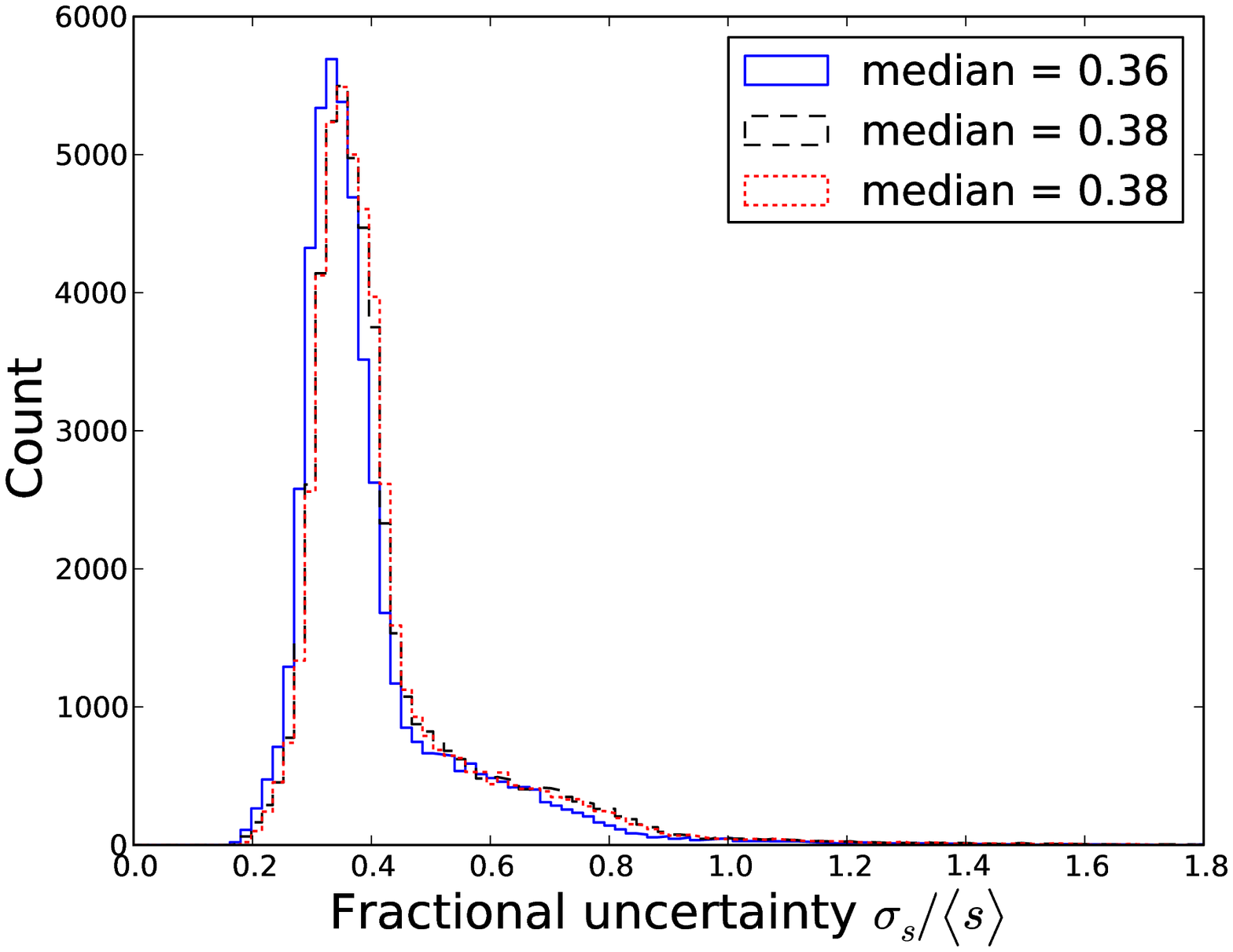}
  \caption{As Fig.~\ref{fig:studentized}, but using two flawed priors as
    described in the text. Blue line, full: correct prior; Black line, dashed:
    approximate prior 1; Red line, dotted: approximate prior 2.}
  \label{fig:flawedresults}
\end{figure}

Fig.~\ref{fig:flawedresults} shows the results of the analysis of our
pseudodata using each of these priors.  The results are remarkably good,
aligning extremely closely with those using the correct prior. This is very
encouraging, as it implies that the use of an approximate prior in the
analysis of a real sample will give very reliable results. Most importantly,
the results in Fig.~\ref{fig:flawedresults} are incontrovertibly better than
those of Fig.~\ref{fig:maxlik}, signalling that the use of an approximate
prior in any photometric distance determination \textit{must} be preferred to
the use of a flat prior -- maximum-likelihood techniques are sub-optimal in
such a complex situation.

\begin{figure}
  \includegraphics[width=84mm]{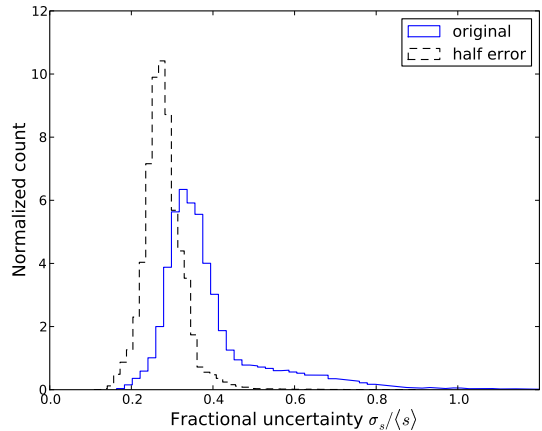}
  \caption{Normalized distribution of quoted fractional uncertainties from
analysis of both the original pseudodata, with full $\sigma_{\log g}$
(`original') and data with half of this gravity error (`half
error').\label{fig:errors2}} 
 \end{figure}

Fig.~\ref{fig:errors2} displays the uncertainty histogram produced from
analysis of a sample generated in exactly the same way as before, except with
the observational error on $\log g$ halved to $\sigma_{\log
g}(\bar\mathbf{y}_1) = 0.5 \,\gamma(\bar\mathbf{y}_1)$\,dex. When analysed
with corresponding uncertainty included, there is a dramatic improvement in
the accuracy obtained (the curve labelled `half error'). This is not
surprising, since surface gravity is the key discriminant in differentiating
between giants and dwarfs, and since the difference in brightness between the
two species is so vast, any improvement in our ability to discriminate
between them will rapidly decrease the uncertainty of our distance estimate.
Quantitatively, halving the error in $\log g$ reduces the uncertainty in the
distance by a factor of order $23\%$.  Hence it is of the utmost importance
to beat down observational errors in measurements of $\log g$ whenever
photometric distances are to be obtained.

The negative wing of the distribution shown in the top panel of
Fig.~\ref{fig:studentized} is largely composed of stars that are best
modelled as old stars ($\tau \geqslant 6$\,Gyr) of around
one solar mass. The underestimation of the uncertainties on such
stars stems from the high weight that the prior assigns to such stars;
if the data can be matched by such a star, the probability of this match is
high and the uncertainty in the distance is dominated by the small error in
the star's apparent magnitude.

\subsection{Metallicities, ages and masses}

\begin{figure}
  \includegraphics[width=84mm]{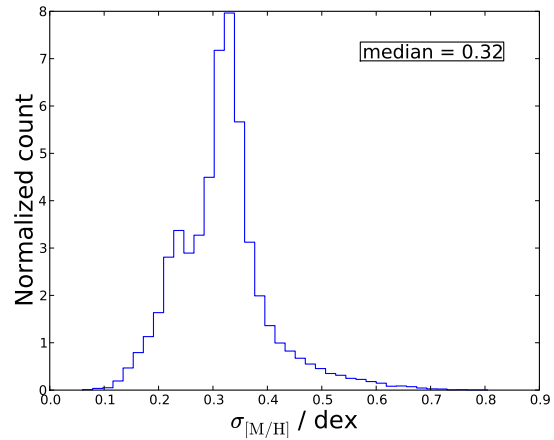}
  \caption{Histogram of the quoted `observational' metallicity errors in our
pseudodata.}
  \label{fig:trueZerrs}  
\end{figure}

\begin{figure}
  \centering
  \includegraphics[width=\thirdwidth]{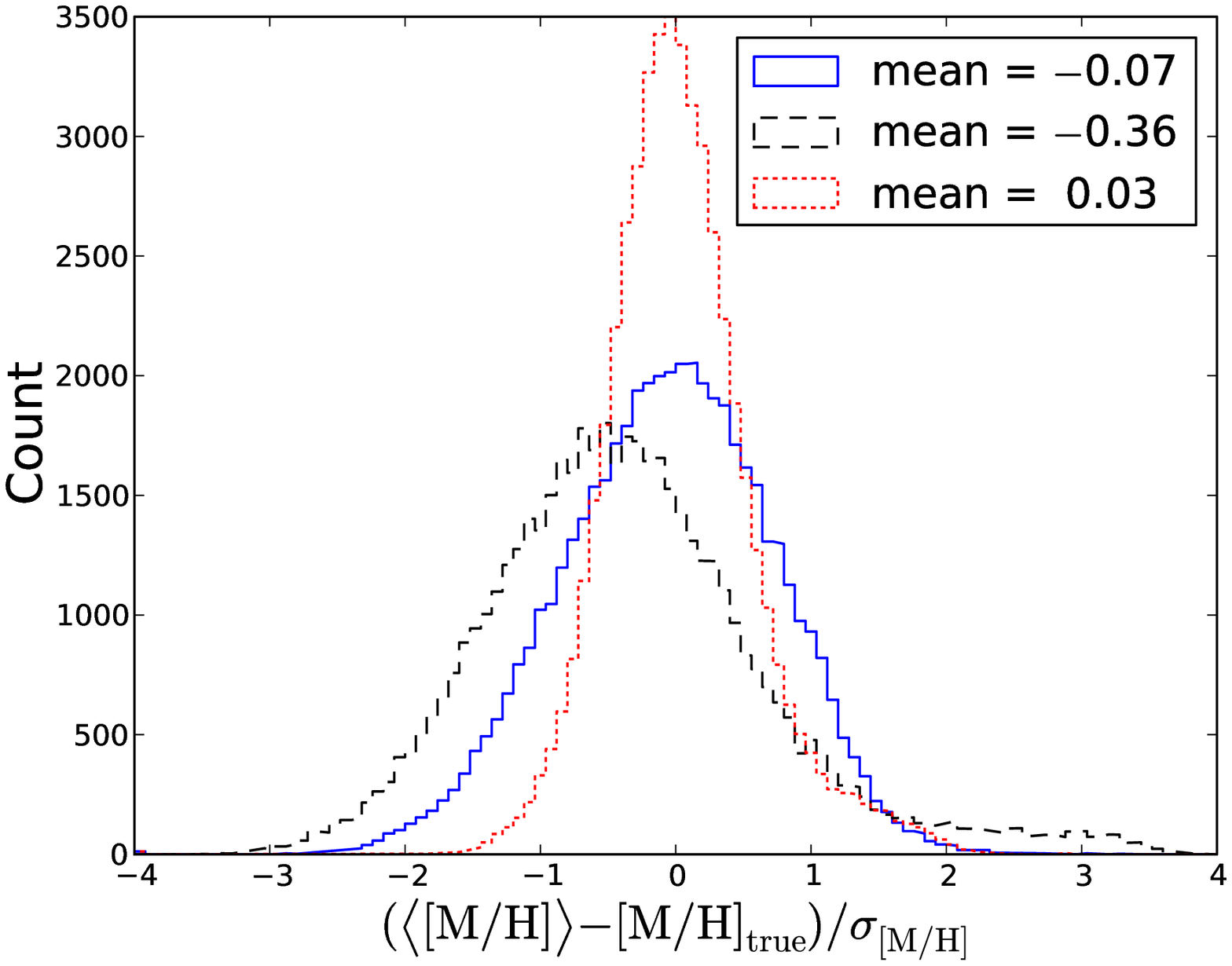}
  \includegraphics[width=\thirdwidth]{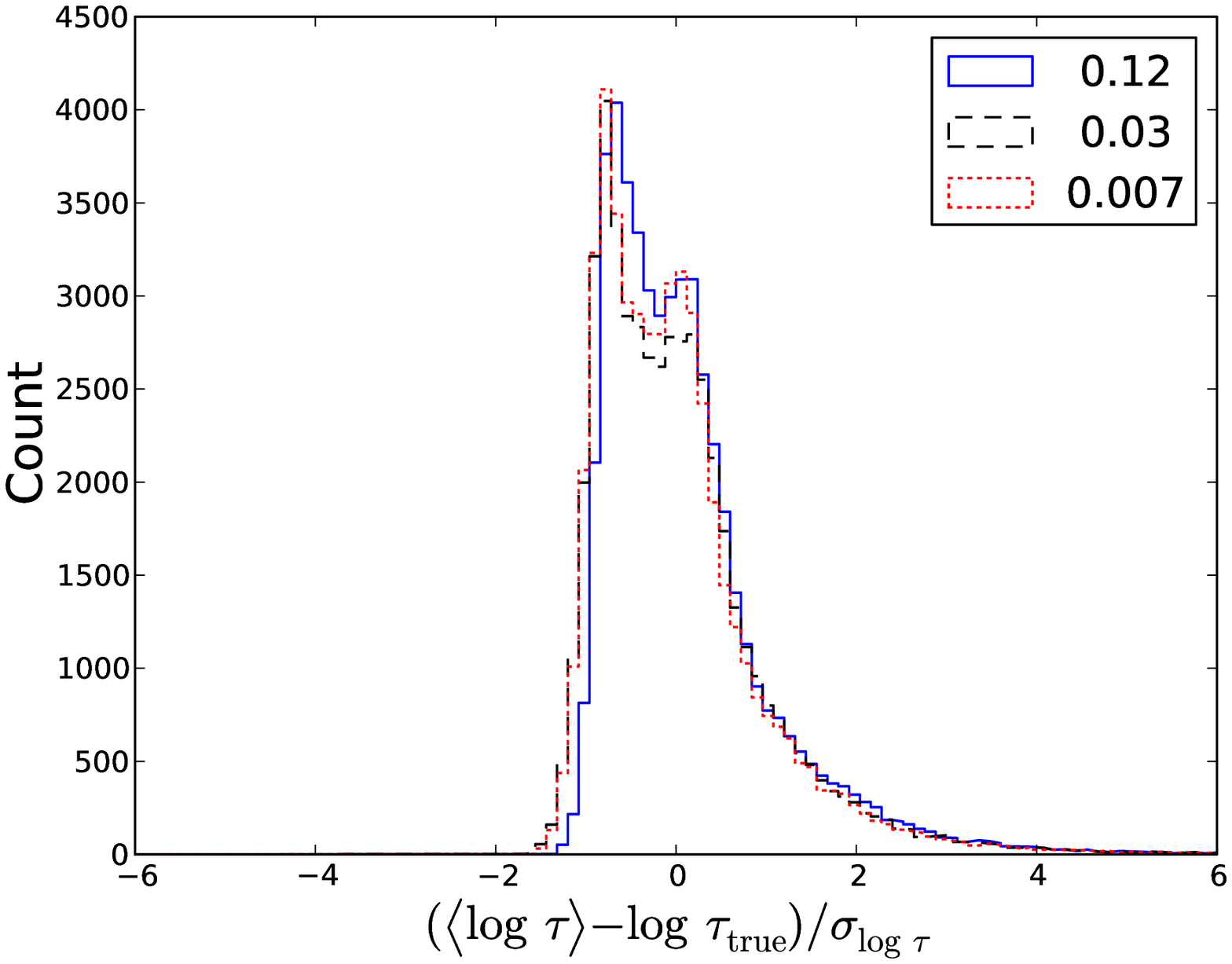}
  \includegraphics[width=\thirdwidth]{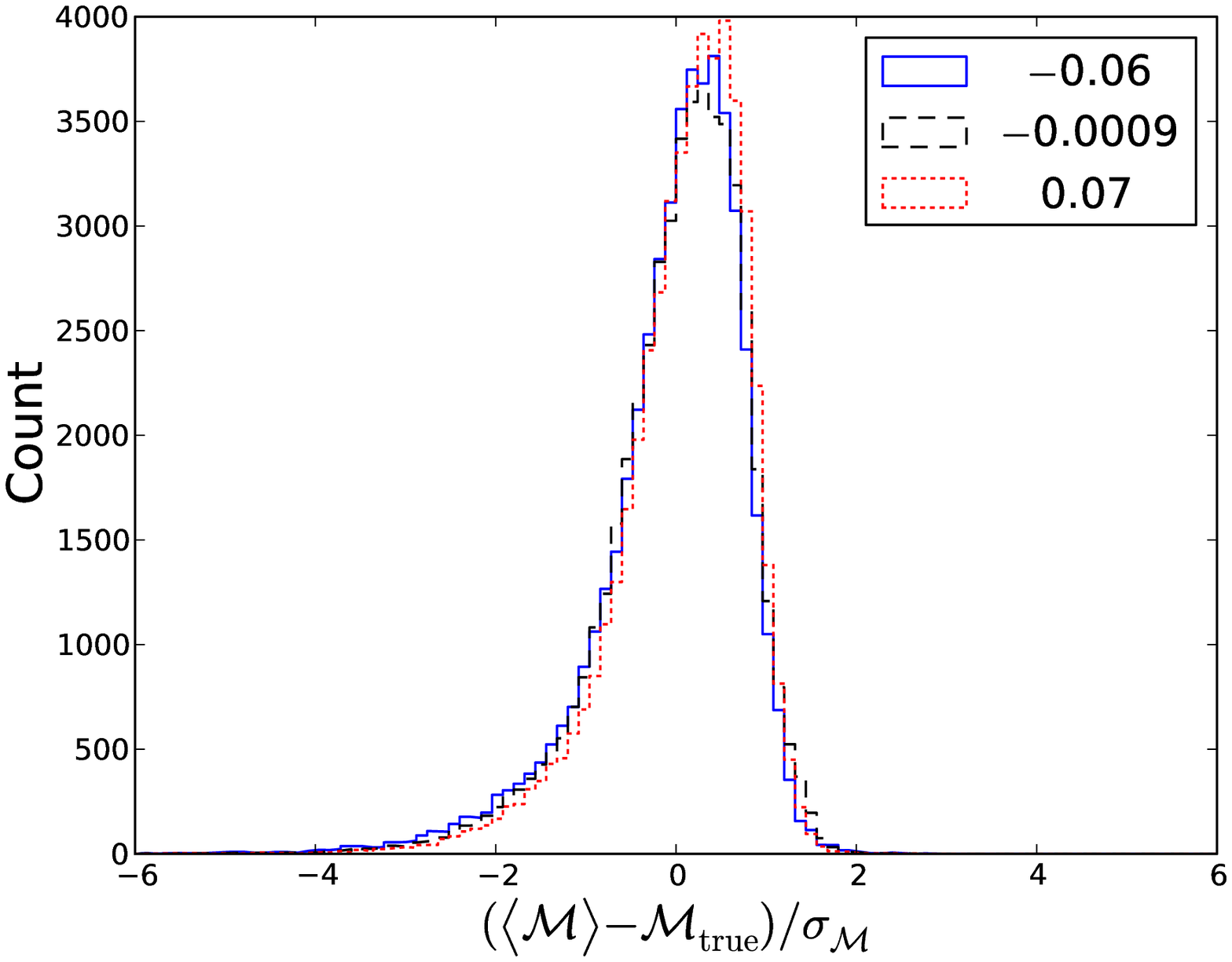}
  \caption{Distributions of residuals in $\mh$, $\log\tau$ and $\cM$ normalized by the corresponding uncertainties 
    for the pseudodata using the correct prior and using the approximate
    priors. Blue line, full: correct prior; Black line, dashed:
    approximate prior 1; Red line, dotted: approximate prior 2. 
    The mean is displayed for each distribution.}
  \label{fig:ztm_hists}
\end{figure}

\begin{figure}
  \centering
  \includegraphics[width=\thirdwidth]{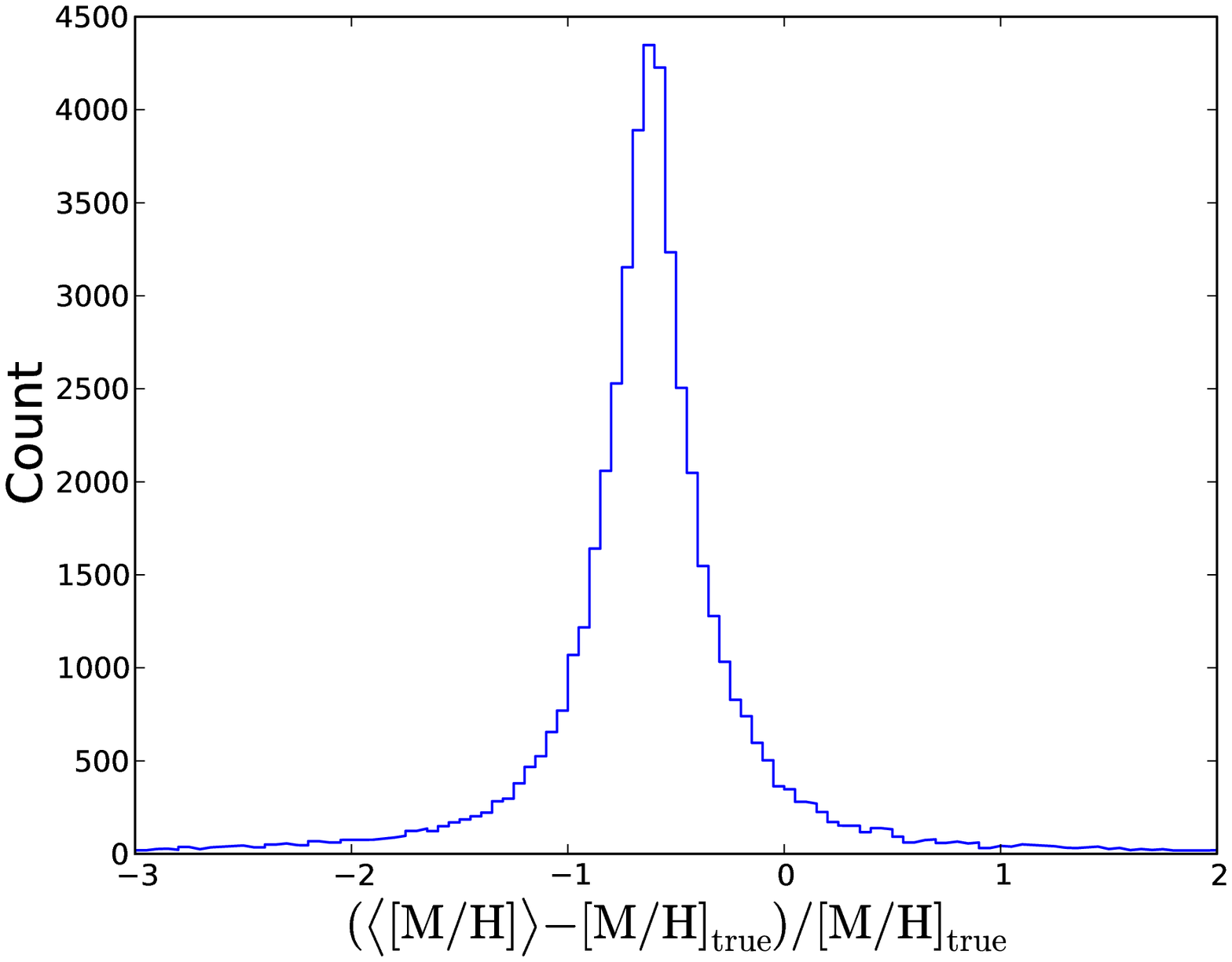}
  \includegraphics[width=\thirdwidth]{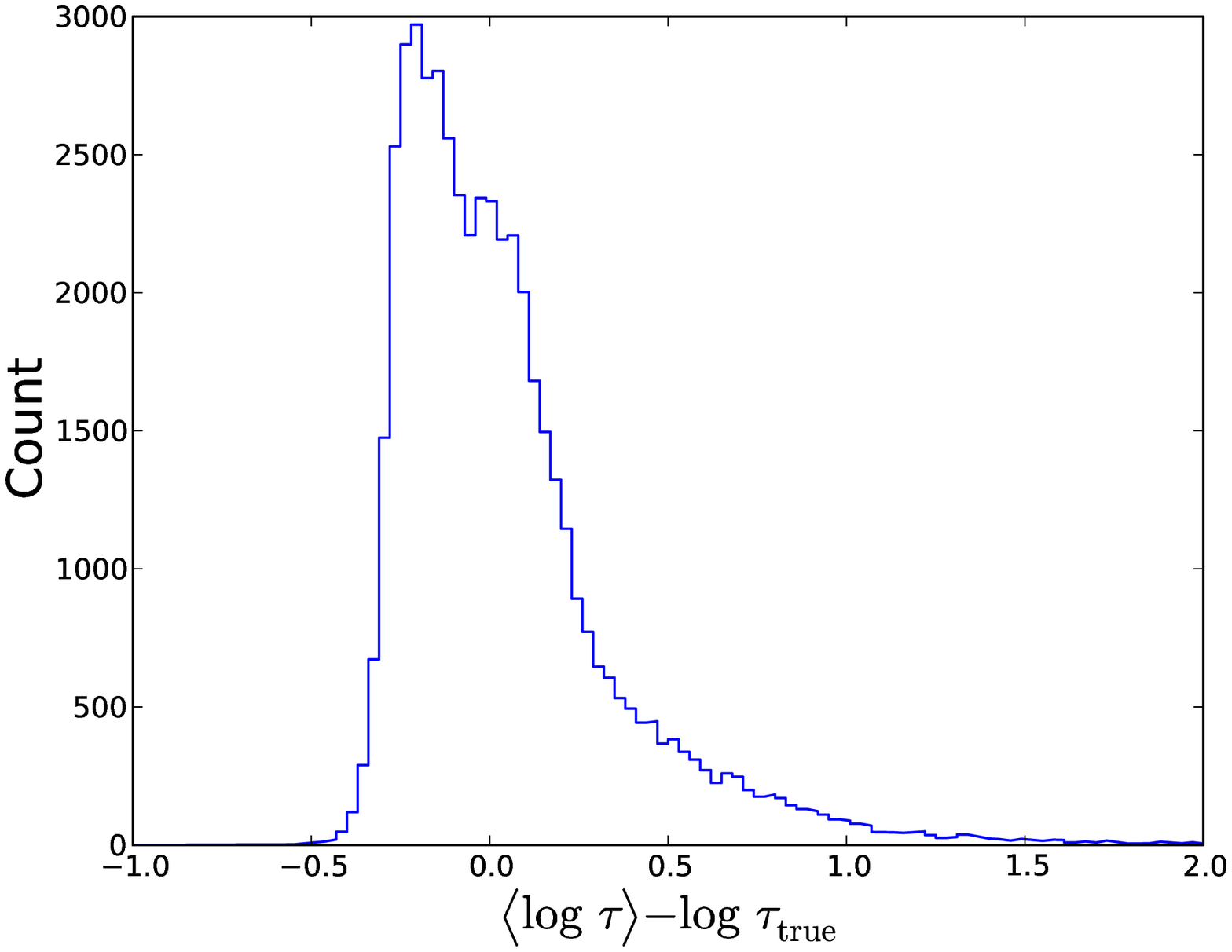}
  \includegraphics[width=\thirdwidth]{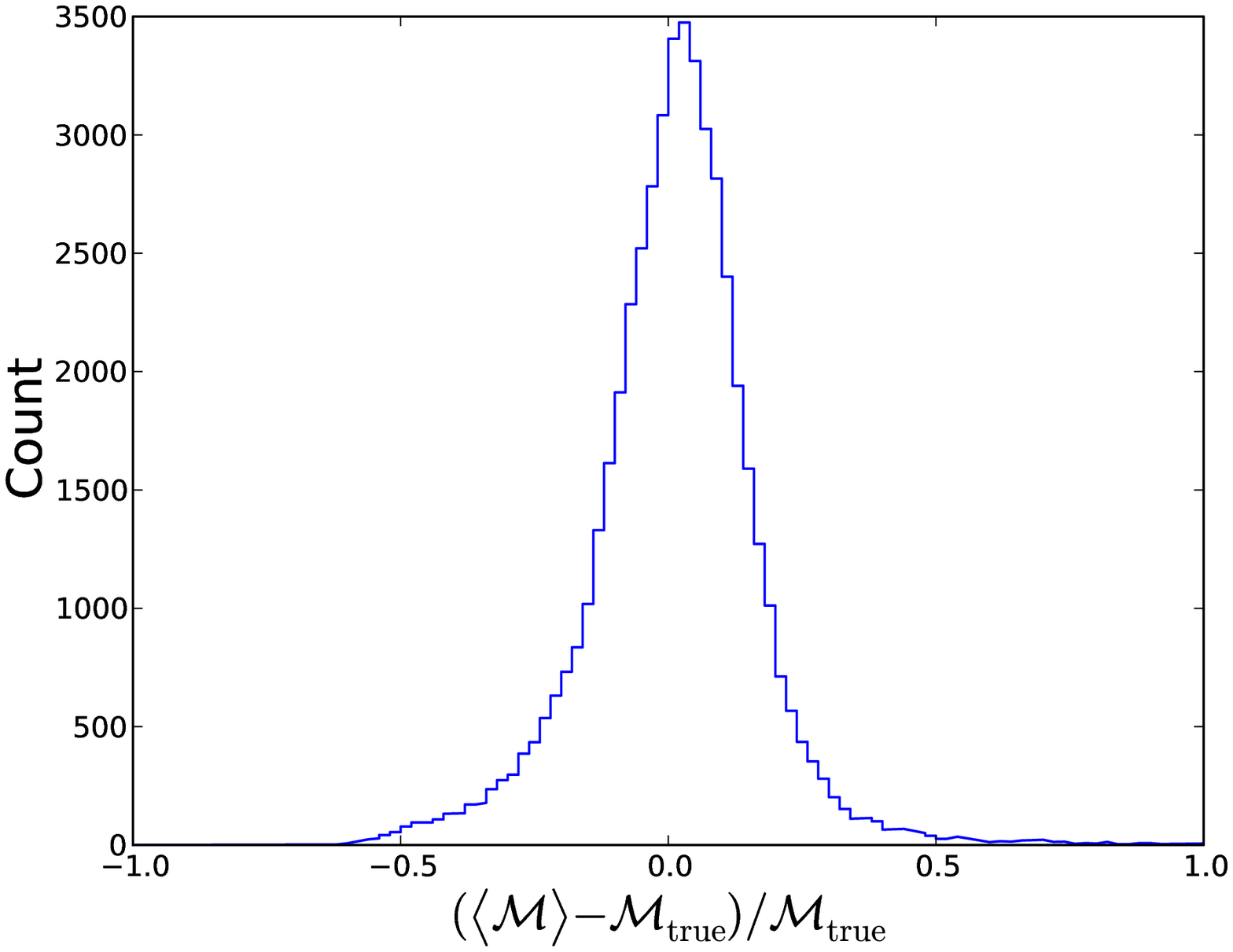}
  \caption{Distributions of residuals (as a fraction of true values in the
    cases of metallicity and mass) in $\mh$, $\log\tau$ and $\cM$, for the correct prior.}
  \label{fig:ztm_resids}
\end{figure}

\begin{figure}
  \centering
  \includegraphics[width=\thirdwidth]{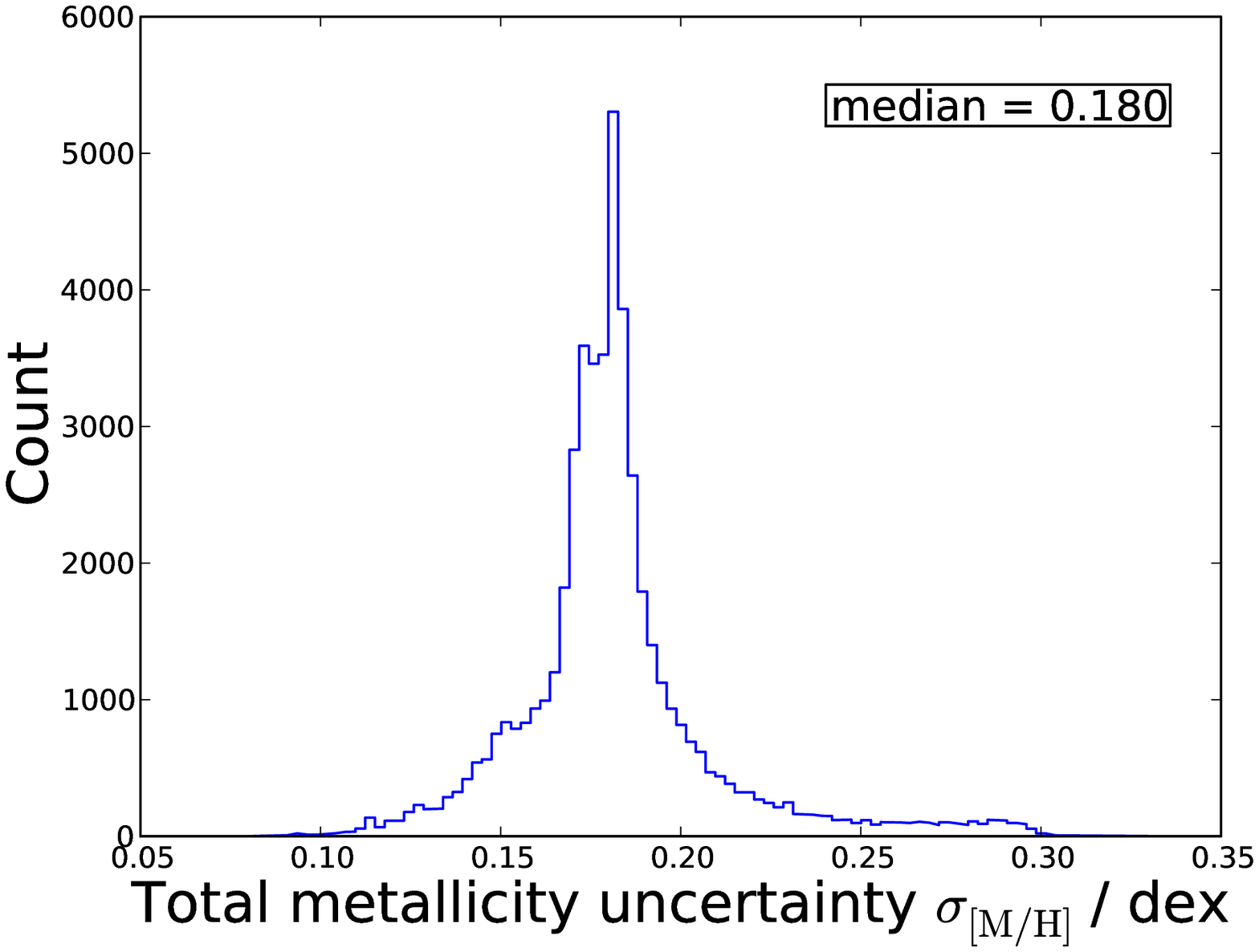}
  \includegraphics[width=\thirdwidth]{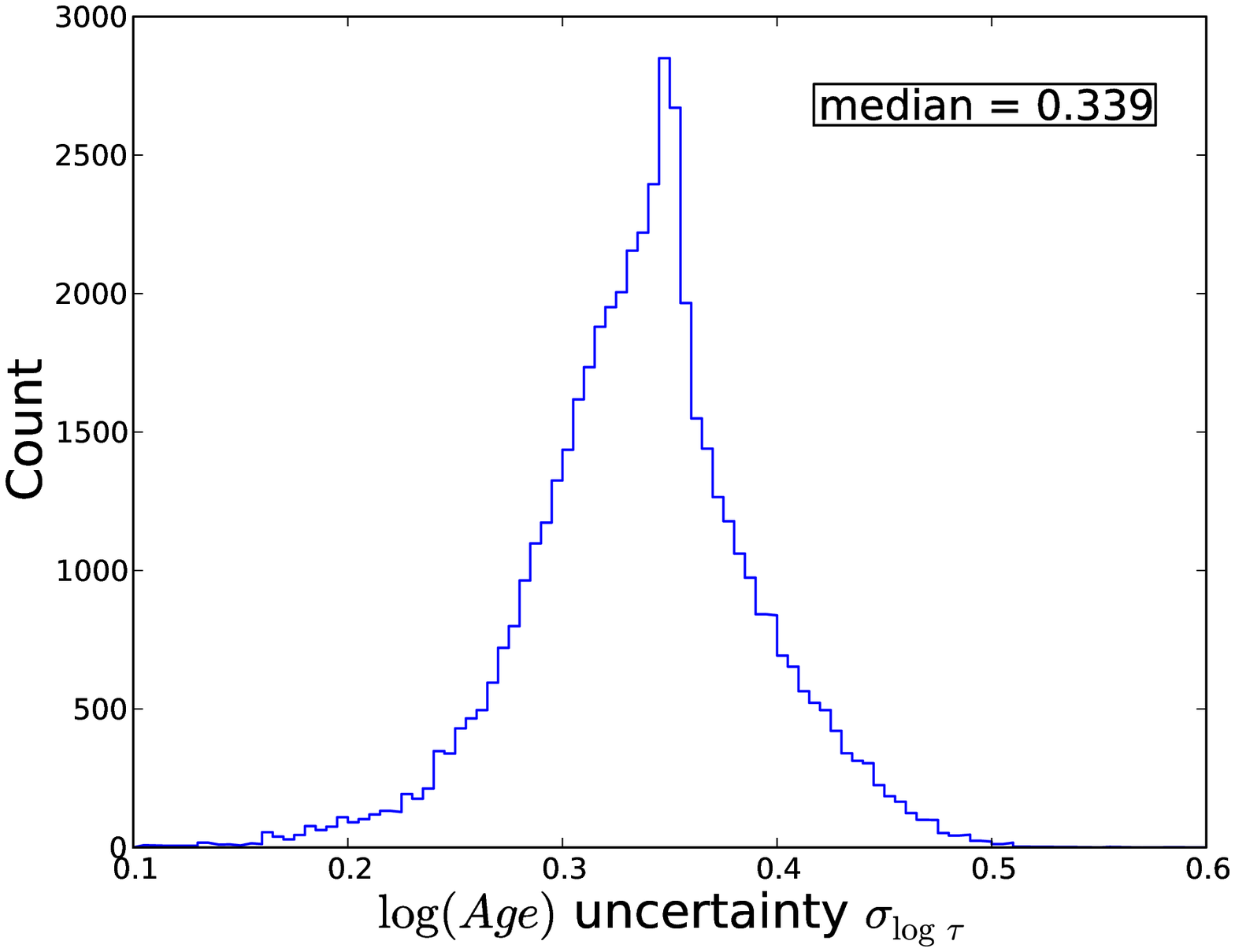}
  \includegraphics[width=\thirdwidth]{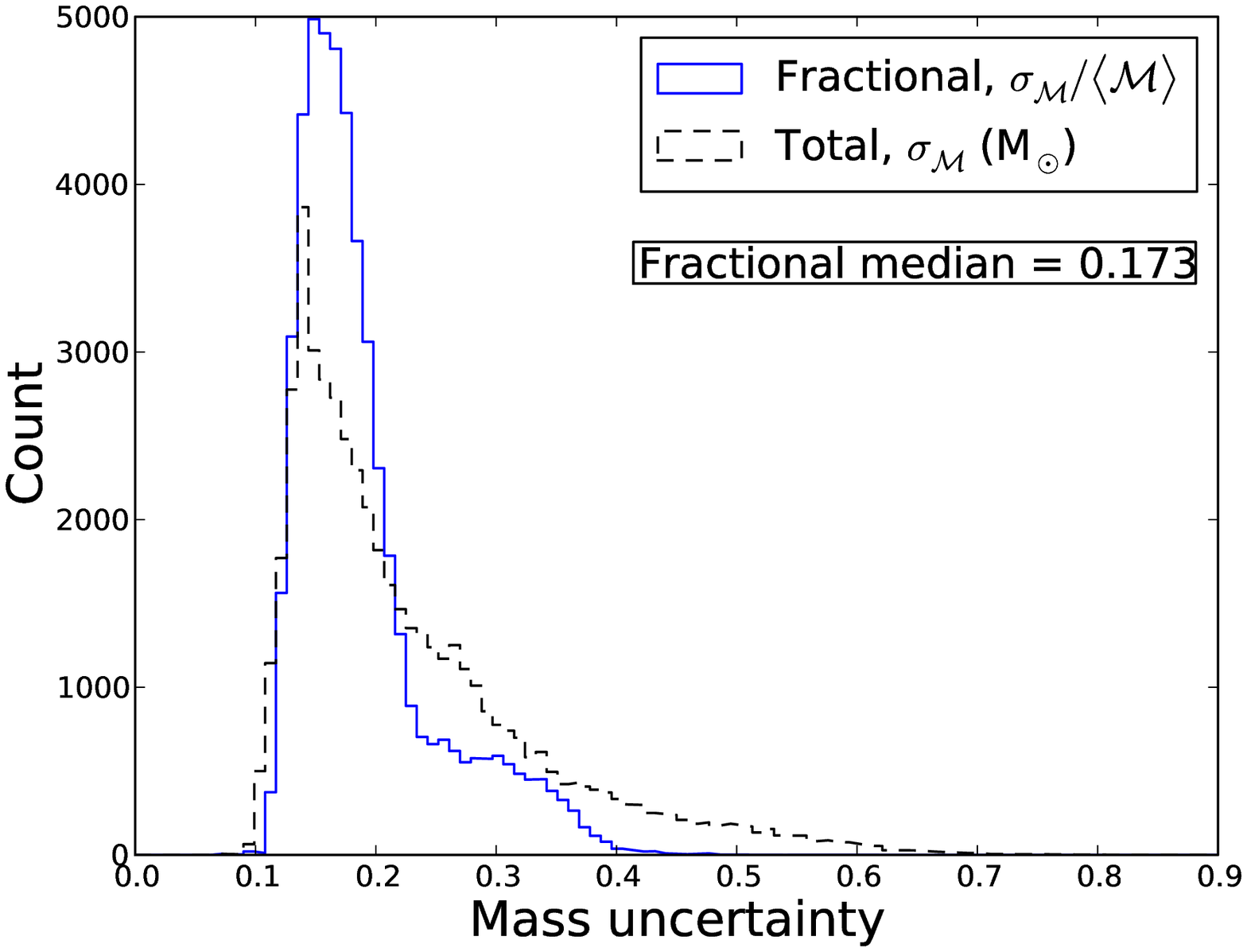}
  \caption{Distributions of uncertainties in $\mh$, $\log\tau$ and $\cM$ for the correct prior.
    Uncertainties in metallicity and log age are absolute; for mass both an
    absolute error (measured in solar masses) and a fractional error distribution
    are displayed.}
  \label{fig:ztm_errs}
\end{figure}

Figs.~\ref{fig:ztm_hists}--\ref{fig:ztm_errs} show the performance of our
method in the recovery of metallicities, ages and masses.
Fig.~\ref{fig:ztm_hists} shows that there is minimal bias in all three
measurements, and that the estimated uncertainties are reliable. Furthermore,
Fig.~\ref{fig:ztm_errs} shows that with estimates of the assumed precision
($0.043\dex$ in $T_{\rm eff}$, $\sim\!0.5\dex$ in $\log g$, $\sim\!0.3\dex$
in $\mh$, $0.045\mag$ in $J-K$ and $0.023\mag$ in $J$) stellar parameters can
be determined with good precision.  Remarkably, the output uncertainties in
$\mh$ are strongly peaked at $\sim\! 0.18\dex$, significantly smaller than
the `observational' errors ($\sim\! 0.32\dex$, see Fig.~\ref{fig:trueZerrs}).
This reduction in uncertainty is made possible by simultaneously using all
available information, which includes the physics of stellar evolution and
the morphology of the Galaxy; it is not in any sense a `creation of
information'. The errors given in the RAVE catalogue are conservatively large
to allow for imperfections in the calibration data sets
(\citealt{RaveDR2_cut}).  They are, moreover, based on the analysis of
individual stellar spectra, without regard to the known properties of the
population from which the individual star is drawn.  It is to be expected
that the injection of prior information about the Galaxy's stellar
populations and the information carried by the photometry diminishes the
uncertainty on the metallicity of each star.

Also included in Fig.~\ref{fig:ztm_hists} are the results obtained with the
poor prior. It can be seen that only the metallicity results are
particularly altered, which is extremely promising: the bad input prior (a
Gaussian centred on the wrong value) could have been dismissed a priori by
comparison with the observed metallicity distribution of the sample, and thus
metallicity space is in fact the least at risk from such effects.
Consequently the biasing seen in the first panel of Fig.~\ref{fig:ztm_hists}
is unrealistically pronounced.

One final note: while extinction and reddening have not been included in this
work, any model of these effects could be included in the analysis simply by
altering the dependence of colours and magnitude on position.

\section{Real-world testing: the Geneva-Copenhagen sample} \label{sec:GCS}

The previous section demonstrated the success of the method applied to a
well-controlled sample of pseudodata. Here we
inspect its performance when applied to real-world data, with all of its
associated noise and limitations. To this end we have analysed the
Geneva-Copenhagen survey data (\citealt{GCS2}), selecting all stars with
Hipparcos parallaxes.

Since error propagation from parallax to distance space is not trivial for
stars with sizeable fractional parallax errors, it is best to
perform the comparison in parallax space itself. Our method can provide an
estimated value and uncertainty for each star's parallax as easily as its
distance and with equal validity, so this comparison is a simple matter.
This also permits comparison with a larger sample of stars than would be
permitted in distance space, since even stars with negative Hipparcos
parallaxes can be used in a parallax-space comparison. Consequently the
subtle biasing introduced by clipping the sample at some parallax or parallax
error can be avoided.

In Section~\ref{sec:pseudodata} we used rather a basic characterization of
the metallicity distribution of the thin disc. The 
Geneva-Copenhagen survey (hereafter GCS) is dominated by the thin disc, so it
is worthwhile to use the best available model of the thin disc's metallicity
distribution.  The analysis of the GCS by \cite{Aumer} found the
underlying metallicity distribution to be well fitted by a pdf of the form 
 \begin{equation}
  p_1([{\rm Fe/H}]) = G([{\rm Fe/H}]+0.12, 0.17),
\end{equation}
 and we used this to represent the metallicity
distribution of the thin disc in equation~(\ref{eq:thindisc}), approximating
$\mh \approx [{\rm Fe/H}]$.

In order to be able to use the Padova isochrones for our analysis of the GCS
data, we required magnitudes in the Johnson system for the stars. Since Tycho
magnitudes are available for the vast majority of the sample
(\citealt{Hipparcos}), we used the transformation relations of \cite{Bessell}
to convert these to Johnson magnitudes, assuming a conservative spread of
0.02\,mag about the transformation lines.

\begin{figure}
  \centering
  \includegraphics[width=\thirdwidth]{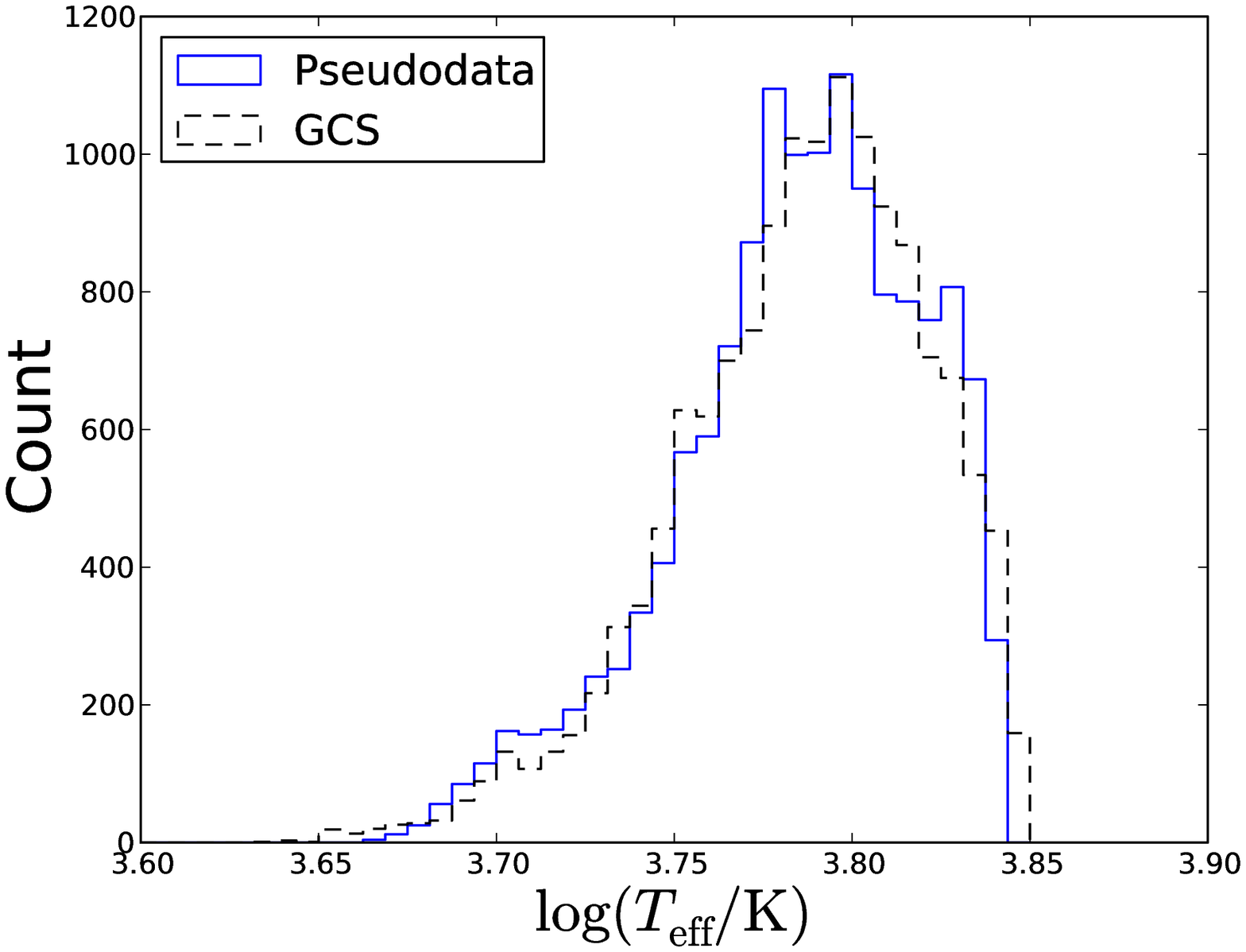}
  \includegraphics[width=\thirdwidth]{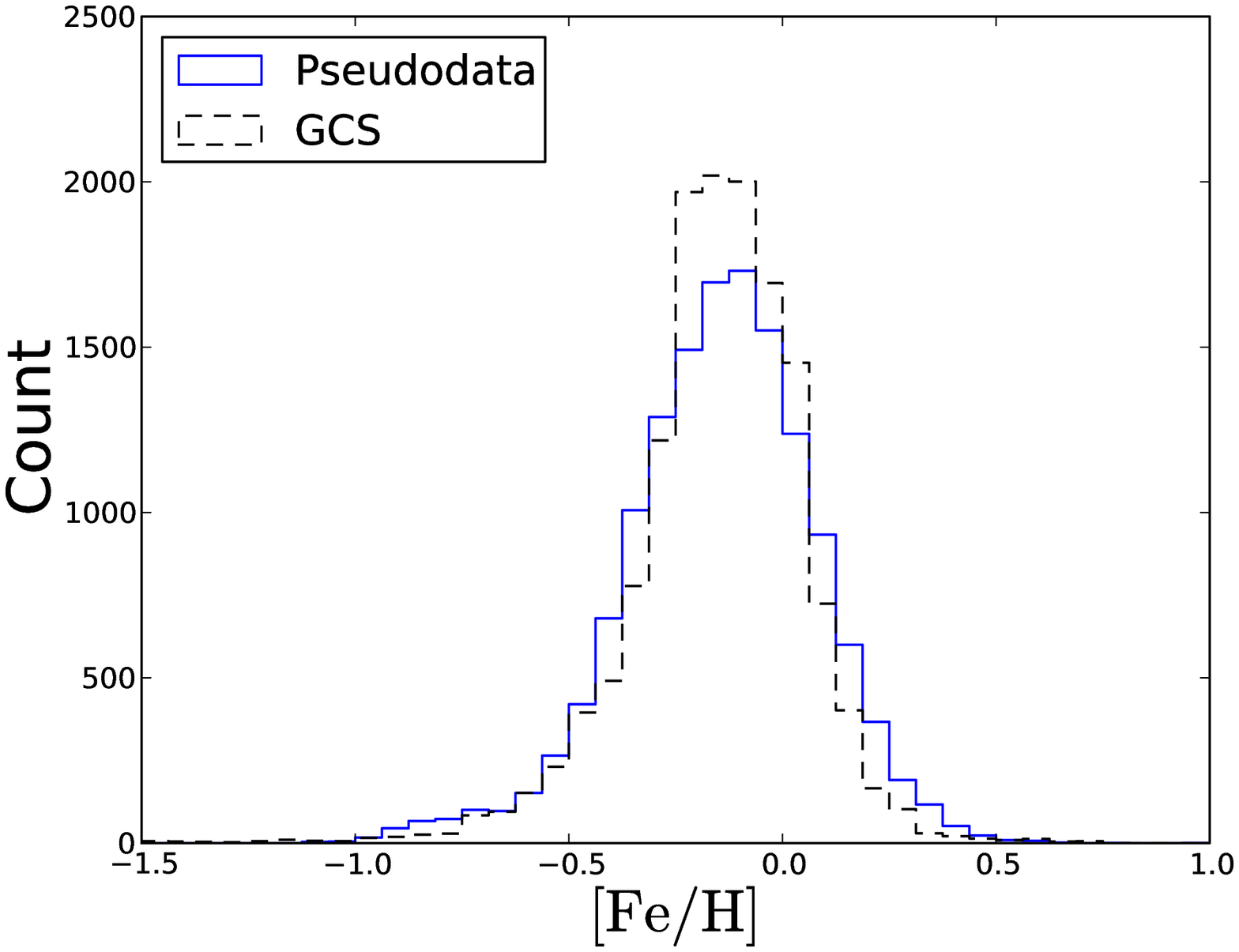}
  \includegraphics[width=\thirdwidth]{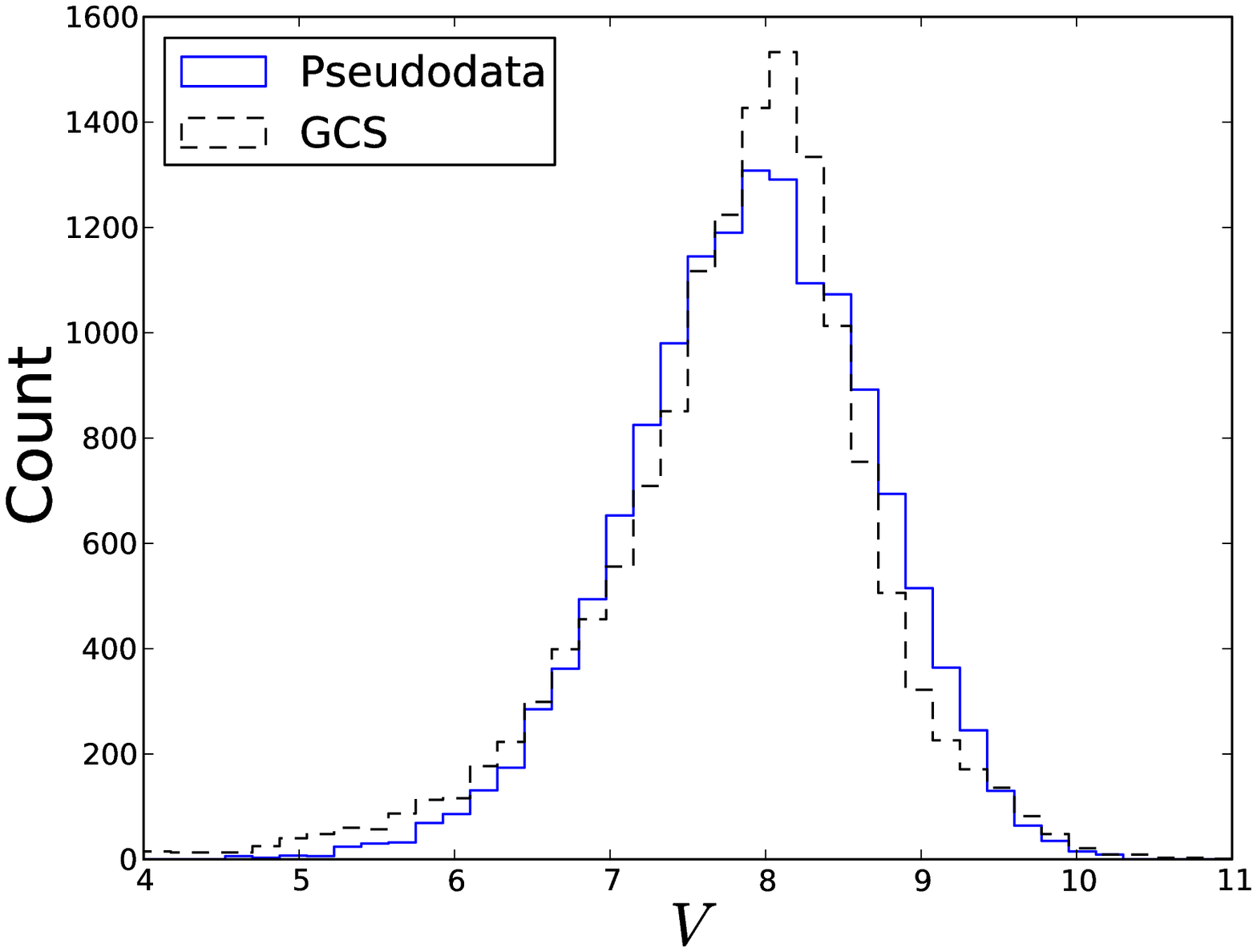}
  \caption{Distribution in effective temperature, metallicity and $V$-magnitude for our GCS pseudodata (full lines) and star in the real GCS catalogue (dashed lines).}
  \label{fig:GCS_match} 
\end{figure}

We then generated a pseudodata set to mimic the GCS. For this set, we
performed the same procedure as in Section~\ref{sec:pseudodata}, taking
 \begin{equation}
  \mathbf{y} = \left( \log T_{\rm{eff}}, V, B-V, \mh_{\rm{obs}} \right),
\end{equation}
 with the selection function defined by the intersection of the following
conditions:
{\arraycolsep=2pt \begin{eqnarray}
   \selfn & = & \cases{1&if $0.28 \leqslant V-J < 1.72$, \cr
                       0&otherwise;} \label{eq:VmJcond}\\
   \selfn & = & \cases{1&if $\log g \geqslant 3$, \cr
                       0&otherwise;} \\
   \selfn & = & \cases{\frac{\log \overline{T_{\rm eff}} - 3.6}{0.24}&if $3.6 \leqslant \log \overline{T_{\rm eff}} < 3.84$, \cr
                       0&otherwise;} \\
   \selfn & = &G(\bar{V}^2 -50,13) .
\end{eqnarray}
}
 Condition~(\ref{eq:VmJcond}) is appropriate because the GCS was designed to
limit the survey to F and G stars; the colour cut we have imposed is
conservative, permitting stellar types in the range A5--K0
(\citealt{GalAst}). Our $\log g$ cut crudely reflects the selection against
giants that was made on the basis of Str\"{o}mgren photometry; the other
factors are simply based on visual inspection to provide a reasonable mimic
of the actual distributions of the data in observable space. We also imposed a 
Gaussian error distribution on the `measured' parallax with dispersion
\begin{equation}
  \sigma_\varpi = \cases{
    0.3\,{\rm mas} & if $\bar{V}<5$,\cr
    (0.3 + 0.06 (\bar{V}-5)^2 )\,{\rm mas} & otherwise,}
\end{equation}
based on an approximate fit to the variation of Hipparcos' parallax error with apparent $V$-magnitude.
The resultant match is displayed in Fig.~\ref{fig:GCS_match}.

For the analysis of the GCS data it was decided to take
\begin{equation}\label{eq:GCS_phi}
  \phi(\mathbf{x}) = \cases{1 &if $0.28 \leqslant V-J < 1.72$, \cr
                            0 &otherwise,}
\end{equation}
 and the prior $p(\mathbf{x})$ from equation (\ref{eq:priorofx}) ff.
Including a $\log g$ dependency within $\phi(\mathbf{x})$ would have been
possible, but it was felt to be too haphazard to simply cut away all stars
with e.g.\ $\log g < 3$, since the GCS selection was made on the basis of
Str\"omgren colours and not on direct  measurements of surface gravity.
Consequently we analysed both the pseudodata and the real GCS data using a
selection function as in equation~(\ref{eq:GCS_phi}).
The fitting was performed in the variables $(\log T_{\rm{eff}}, V, B-V)$,
since we were not confident in the precision of matching $\mh$ to $[{\rm
Fe/H}]$.

\begin{figure}
  \centering
  \includegraphics[width=\thirdwidth]{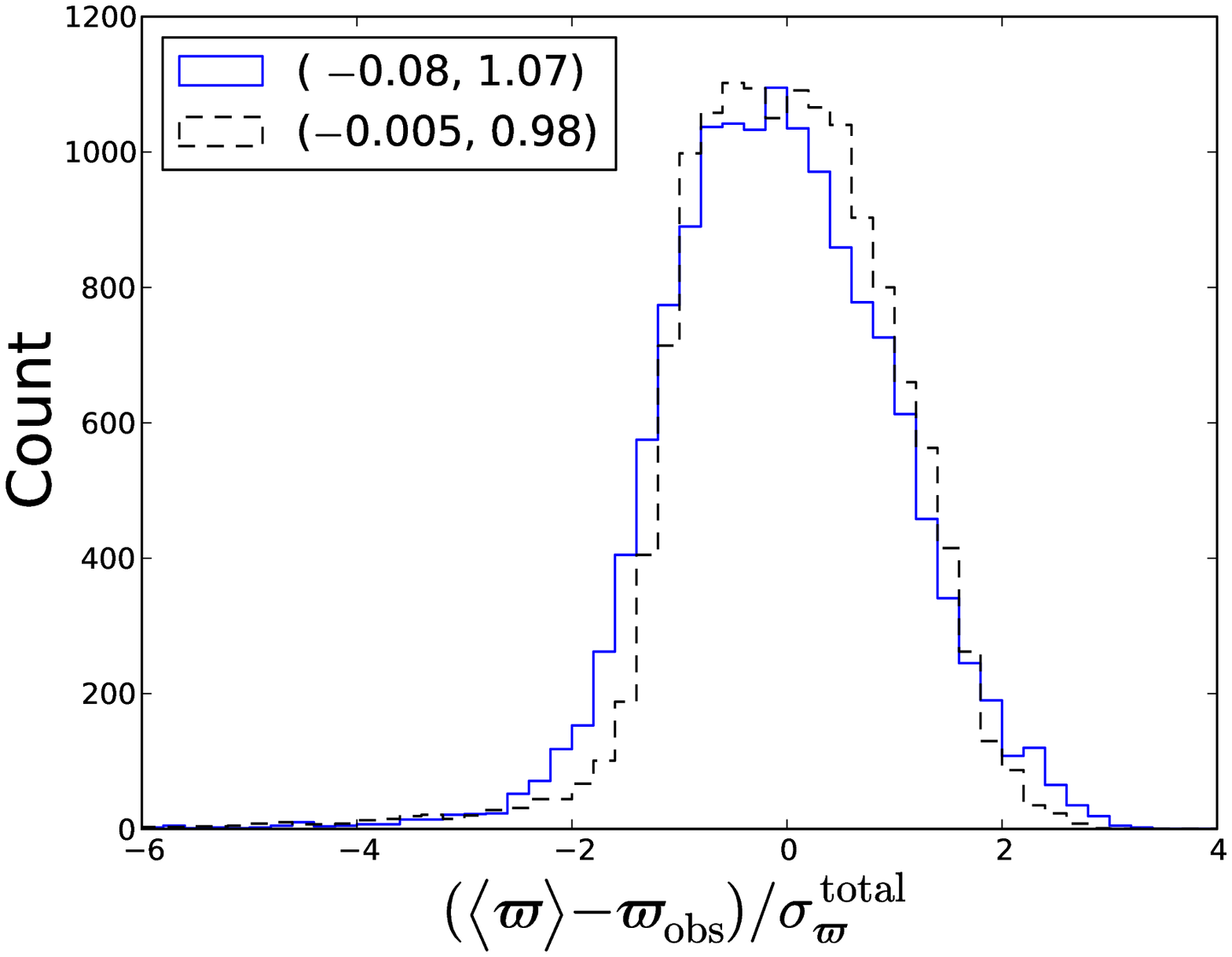}
  \includegraphics[width=\thirdwidth]{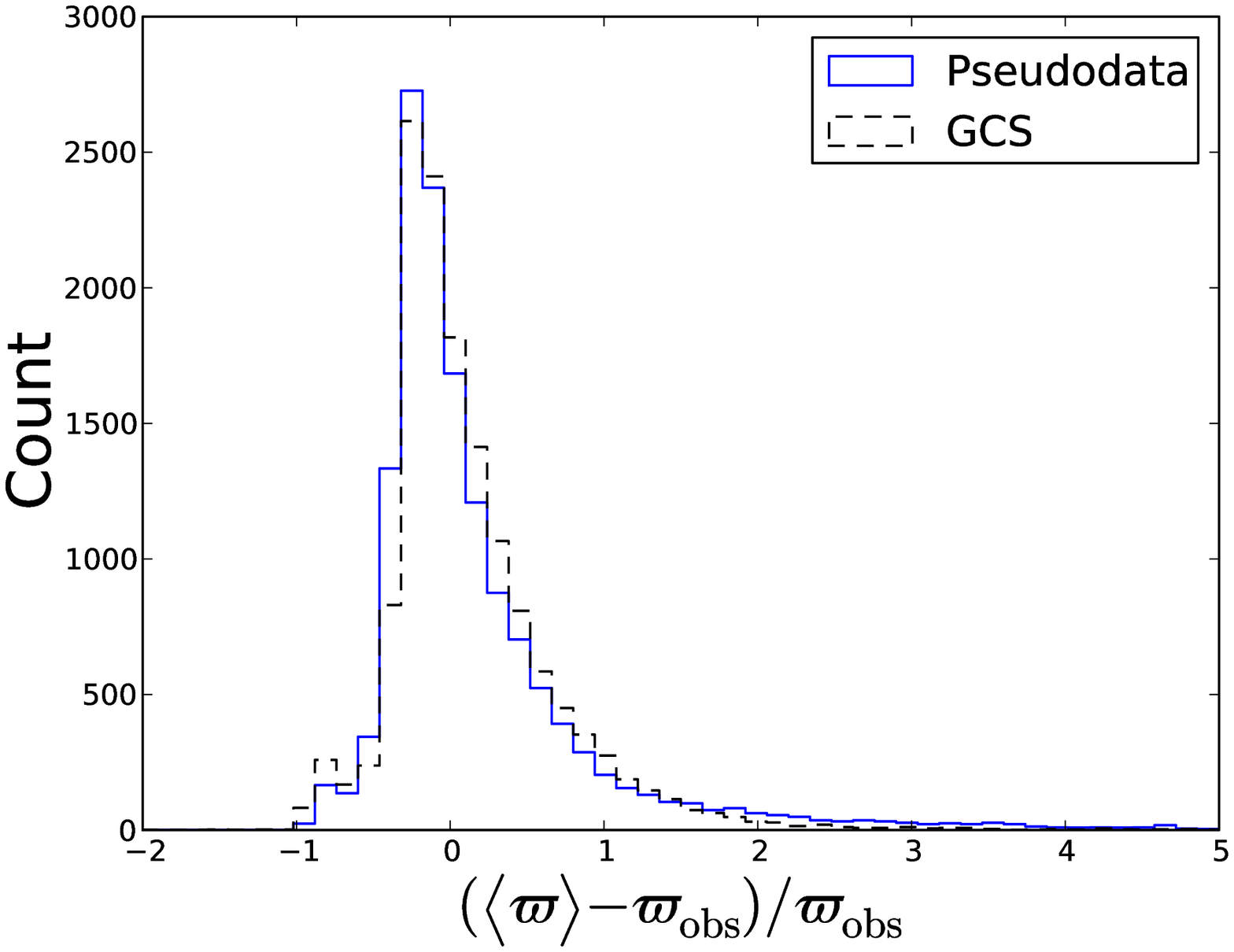}
  \includegraphics[width=\thirdwidth]{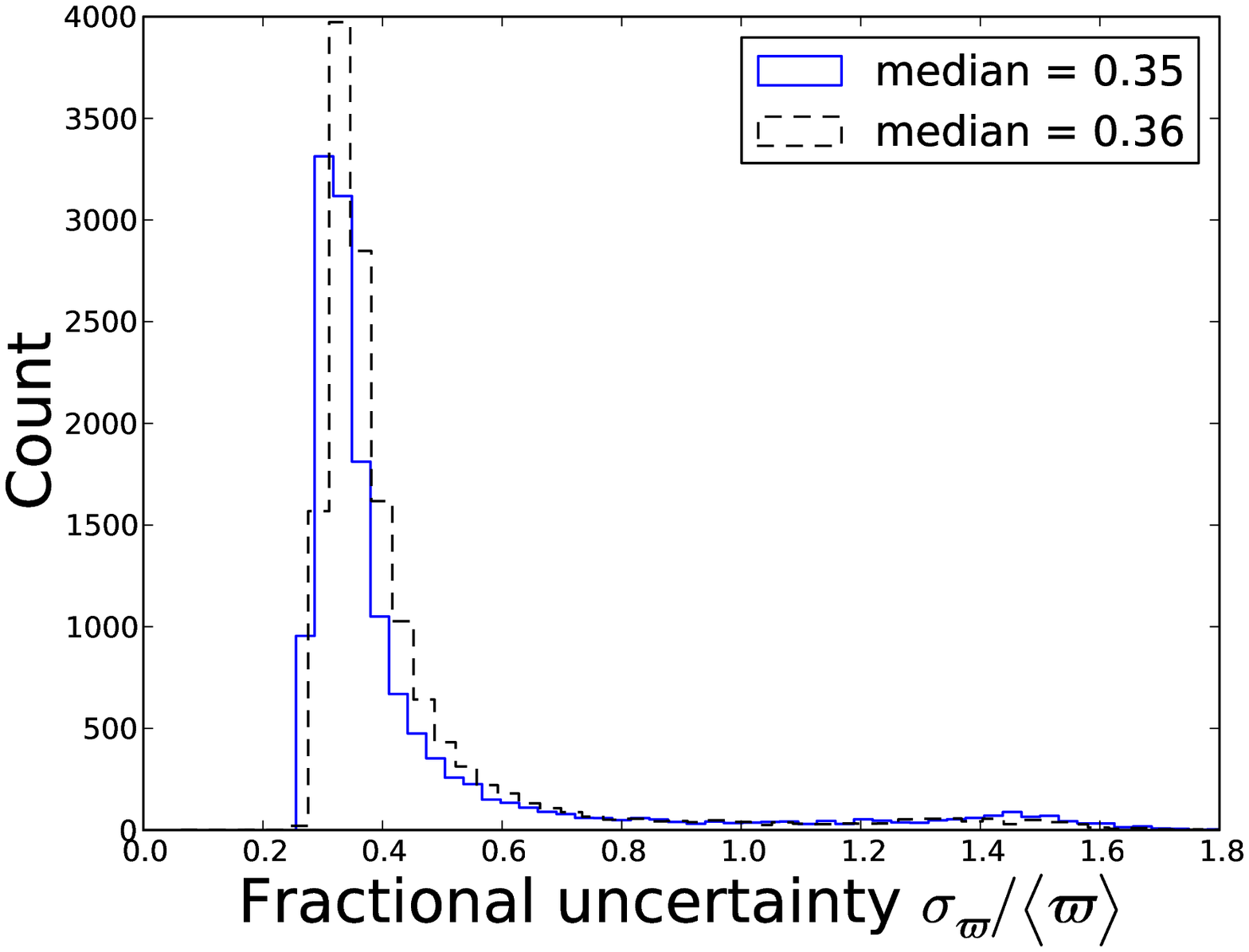}
  \includegraphics[width=\thirdwidth]{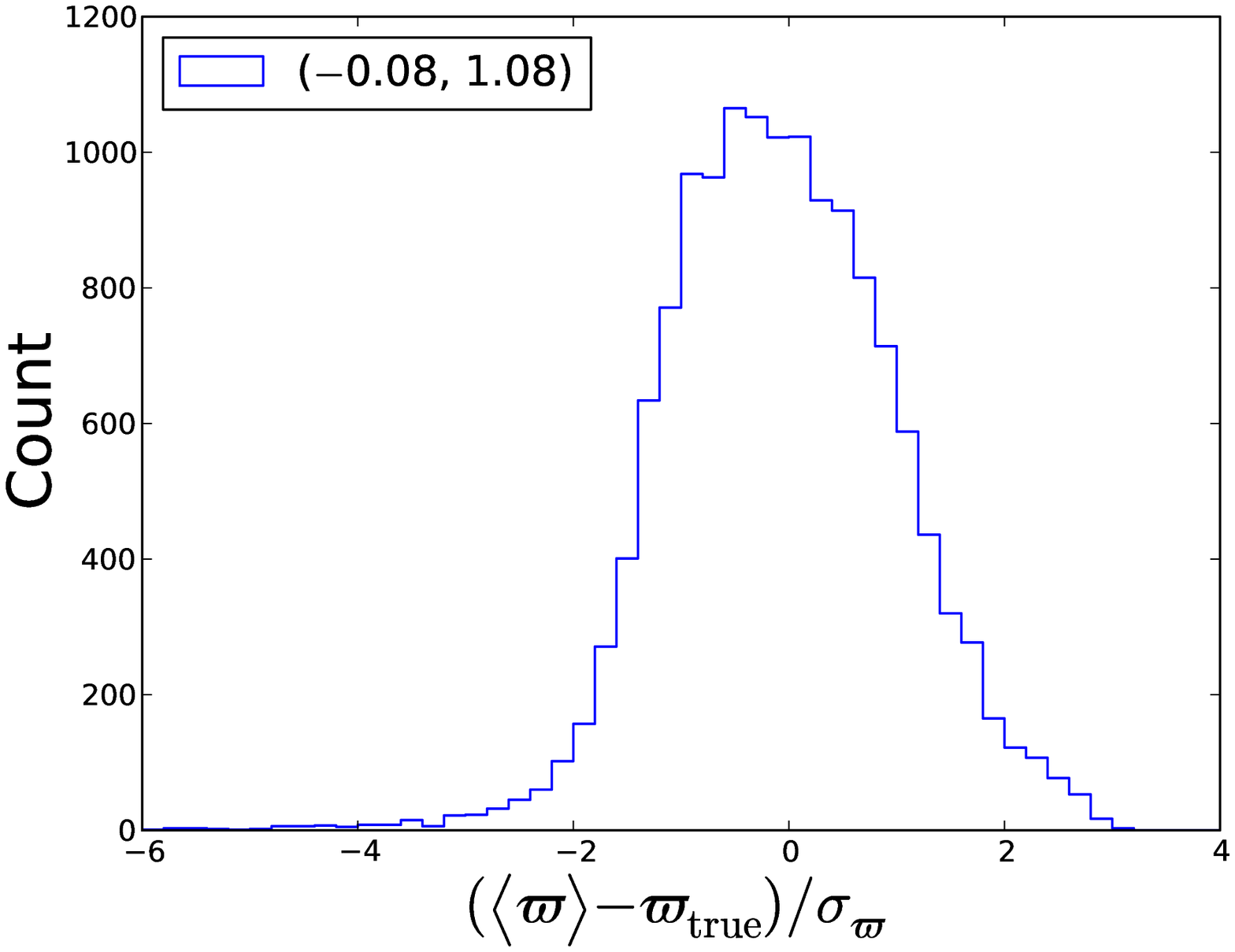}
  \caption{Photometric parallaxes from GCS pseudodata (blue lines, full)
    and from the real GCS sample (black lines, dashed). Top panel: the normalized
    distribution of parallax residuals; $\varpi_{\rm obs}$ is obtained by adding
    the errors returned by the method and the Hipparcos errors in quadrature.
    Each line is labelled by (mean, dispersion). Second panel: distribution of
    residuals as a fraction of true parallax.  Third panel: distribution of
    fractional photometric (output) uncertainties. Bottom panel: equivalent to the top panel but showing 
    true normalized residuals for the pseudodata, removing the `observational' error.}
  \label{fig:GCS_results}  
\end{figure}

\begin{figure}
  \centering
  \includegraphics[width=84mm]{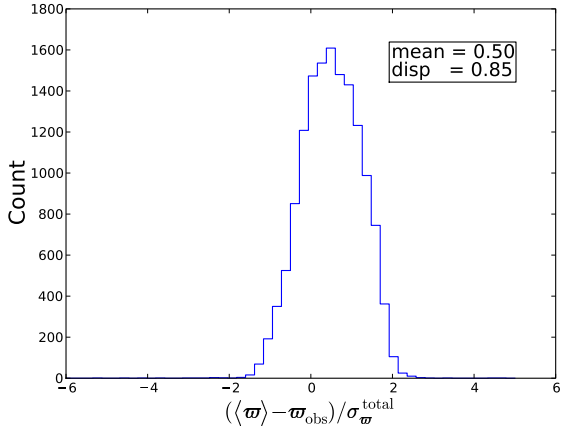}
  \caption{Normalized parallax residuals for the real GCS sample for a maximum likelihood analysis.}
  \label{fig:GCS_ML}  
\end{figure}

In Fig.~\ref{fig:GCS_results} we display the results of analysis of both the
pseudodata and the true GCS sample; both samples contained 14\,233 stars. It
can be seen that the match is very good. In the first and fourth panels we display the
mean and standard deviation of each distribution. To find these values for
our analysis of the true data we have clipped 62 outliers with
$x$-axis values below $-6$. The total uncertainty in the first panel,
$\sigma_\varpi^{\rm total}$, is found from adding our output uncertainty and
the Hipparcos error in quadrature. The uncertainties in the third and fourth panels are those
purely due to our method; the fourth panel displays the normalized residuals for the pseudodata analysis when one does not include any observational error in the `observed' parallax measurements. It is clear from Fig.~\ref{fig:GCS_results} that our method works
convincingly on real data. Fig.~\ref{fig:GCS_ML} shows the output from applying a maximum likelihood analysis to the same data, and it can be seen that the results are comparable to those from section~\ref{sec:pseudodata}: the mean of the normalized residuals distribution is significantly biased, falling at a value of 0.50.

\subsection{Binary stars}\label{sec:binary}

The method developed here
is properly applicable only to single stars; however binaries with a mass
ratio reasonably far from unity will not present a significant problem since
the observations can be expected to capture the properties only of the more
massive star. It is instructive to consider the worst case -- that of two
equally massive stars -- in which the absolute magnitude will be 0.75\,mag
brighter than for one of the stars. If all other observables are as they
would be for one of the stars, this can be expected to result in an
underestimate of the distance by around 29 per cent.
Fig.~\ref{fig:studentized} demonstrates that this is within, or on the order
of, the output uncertainties for virtually all stars due to the underlying
physics, and thus does not dramatically compromise our results.

\begin{figure}
  \includegraphics[width=84mm]{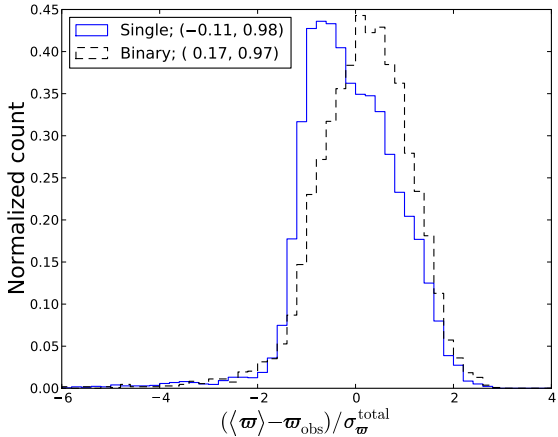}
  \caption{Histogram of the normalized residuals in parallax space for stars
    thought to be single (blue line, full) and those flagged as binaries (black
    line, dashed). For each distribution, the mean and dispersion are displayed
    in the legend.}
  \label{fig:GCS_binaries}  
\end{figure}

The GCS catalogue makes an empirical test of this theory possible.
Fig.~\ref{fig:GCS_binaries} shows the histogram of normalized parallax
residuals for our analysis of the GCS data, divided into those that have been
flagged in the GCS catalogue as a probable binary, and those without this
flag. Each group is labelled with its mean and dispersion after cutting
outliers at $x<-6$. The photometric parallaxes of the binaries are larger
than their measured parallaxes by $0.17\sigma$ in the mean, while the
parallaxes of the single stars are $0.11\sigma$ smaller in mean than their
measured parallaxes.  Since $\sigma/\varpi\sim0.36$
(Fig.~\ref{fig:GCS_results}), this level of bias 
corresponds to $\sim\! 6$ per cent for each star, which
 lies well within the
predicted worst-case level. Moreover, since the probability that a given
system will be flagged as a binary is an increasing function of 
received flux, and therefore of \textit{true} parallax, flagged stars may
be expected to have Hipparcos parallaxes that are on average slightly smaller 
than their true parallaxes.
This can be seen by considering an imaginary sample of stars with a single measured Hipparcos
parallax; these stars will have been scattered by observational errors from true
parallaxes both above and below the measured value. If one then selects, independently, stars with 
a preferentially larger true parallax, one will obtain a sample which is fundamentally
biased to true parallaxes greater than the Hipparcos measurements. Expanding this
reasoning to the entire Hipparcos sample, it must hold for every measured parallax, and thus
one expects the subsample flagged as binaries to be biased in exactly this manner.
Consequently the amount by
which our photometric parallaxes exceed the true parallaxes is likely to be
smaller than the amount by which they exceed the Hipparcos parallaxes. In
fact this effect may also account for the discrepancy between the photometric and
Hipparcos parallaxes of `single' stars: many of these objects will be binary
stars that are too distant to be resolved, and it is likely that in more than
half of these cases accidental errors have boosted rather than
diminished the Hipparcos parallax.

\subsection{Parallax vs distance}

\begin{figure}
  \includegraphics[width=84mm]{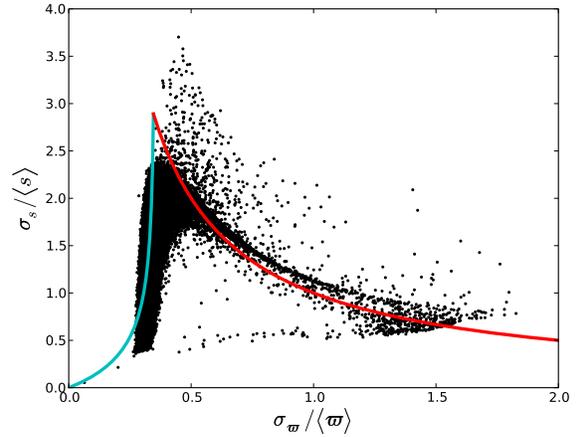}
  \caption{The distribution of stars in fractional error space, from analysis
of the GCS data. Fit curves are described in Section~\ref{sec:GCS}.}
  \label{fig:GCS_stiletto}  
\end{figure}

It is of interest to inspect the results of our analysis of the GCS in
distance as well as in parallax. Fig.~\ref{fig:GCS_stiletto} shows a
plot of the distribution of output fractional uncertainties in distance
against fractional uncertainties in parallax. It follows a rather peculiar
`high-heeled shoe' shape. We can explain the general form of this
distribution by considering two highly simplified forms for a star's distance
pdf. Inspection of the pdf for stars in the `heel' of the shoe
($\sigma_\varpi / \langle\varpi\rangle \sim 0.3$) shows the typical pdf to be
reasonably approximated by a single power-law with a low-$s$ cutoff, i.e.
 \begin{equation}
  p(s) \propto \cases{0 & for $s<a$,\cr
                      s^{-\alpha} & otherwise.}
\end{equation}
 If we calculate the moments of this highly simplified pdf we find that,
letting $X \equiv \sigma_\varpi / \langle\varpi\rangle$ and $Y \equiv
\sigma_s / \langle s \rangle$, we should expect the relation
 \begin{equation}
  \sqrt{1+X^{-2}} + \sqrt{1+Y^{-2}} = 2 ,
\end{equation}
 which is overplotted on Fig.~\ref{fig:GCS_stiletto} and can indeed be seen
to fit the heel of the shoe.

Regarding stars on the `sole' of the shoe, inspection of their pdfs in
distance reveals them to be largely bimodal. As a very crude approximation of
this we can take
 \begin{equation}
  p(s) = \beta \,\delta\left(s-s_1 \right) + \left(1-\beta\right) \delta\left(s- \eta s_1 \right),
\end{equation}
 where $\delta(x)$ represents the Dirac delta function. The resultant moments
are not trivial, but if we consider only the case in which $\eta \gg 1$, we
obtain
 \begin{equation}
  (X,Y) \approx \left( \sqrt{\frac{1-\beta}{\beta}} , \sqrt{\frac{\beta}{1-\beta}} \right) \quad \Rightarrow \quad Y \approx \frac{1}{X},
\end{equation}
 which is also overplotted on Fig.~\ref{fig:GCS_stiletto} and provides a good
fit to the sole of the shoe, despite the crudeness of the approximation.

The fact that this extremely simplistic widely-spaced bimodal distribution
provides such a good fit to Fig.~\ref{fig:GCS_stiletto} conceals an important
fact. It is related to the lack of $\log g$ information in the GCS data in
the form that we have used them: the distance-fitting algorithm has essentially no
way of distinguishing between dwarfs and giants for many of the stars,
resulting in two widely-spaced peaks in the distance pdf. The large
uncertainties on the distance reflect this lack of discrimination, providing
an even more cogent argument than that of Fig.~\ref{fig:errors2} for the
necessity of tight observational constraints on $\log g$ if
accurate spectrophotometric distances are to be obtained. In this case it is clear that the
best choice for an indication of each star's position is actually given by
its measured parallax and the uncertainty thereon, as opposed to the direct
distance determination provided by the method.

\begin{figure}
  \includegraphics[width=84mm]{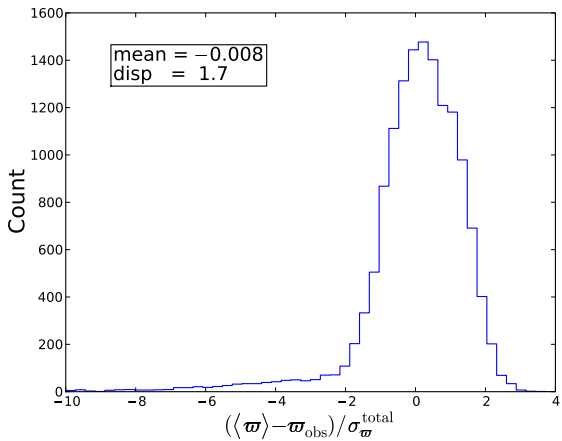}
  \caption{The distribution of normalized residuals from an analysis of the GCS data using the BaSTI stellar models.}
  \label{fig:basti}  
\end{figure}

\begin{figure}
  \includegraphics[width=84mm]{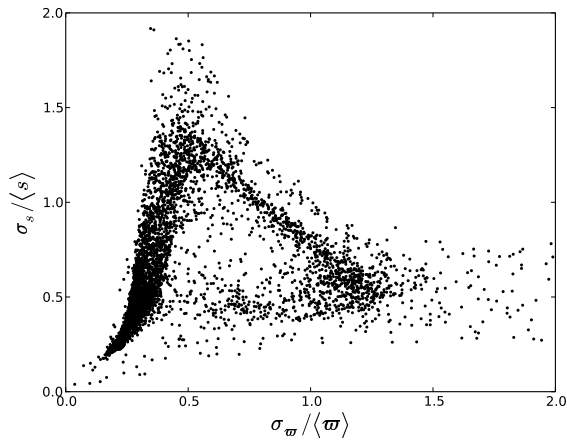}
  \caption{The distribution of stars in fractional error space, from analysis
of the GCS data using the BaSTI stellar models and Str\"omgren photometry.}
  \label{fig:basti_errors}  
\end{figure}

Since the GCS stars virtually all have Str\"omgren photometry, it is of
interest to study the results from running our method when these data are
included. To do this we need a set of stellar models that provide Str\"omgren
magnitudes.  The BaSTI isochrones (\citealt{basti}) provide such models. We
used the $[{\rm Fe/H}]$ values in the set 
 \begin{eqnarray}
  [{\rm Fe/H}]& \in & \{-2.27,-1.79,-1.49,-1.27,-0.96, \nonumber\\
  &&-0.66,-0.35,-0.25,0.06,0.26,0.40\} .\nonumber
\end{eqnarray}
 Unfortunately, it proved impossible to obtain a convincing match between
these models and the data: for many GCS stars ($\sim\! 11\,000$ of the $14\,233$) no BaSTI model lay within
$\sim\! 2\sigma$ of its measured properties -- the errors on the GCS data are
$\sim\! 0.01$ in $\log T_{\rm eff}$ (\citealt{Nordstrom}) and $\sim\!  0.003$
in $b-y$ (\citealt{Olsen1,Olsen2,Olsen3,Olsen4}). 
It is also clear from an inspection of the data that a simple systematic shift in effective temperature cannot solve this problem.
In an attempt to work around this difficulty, we dropped $b-y$ from
$\mathbf{y}$, making the observational constraints
 \begin{equation}
\mathbf{y} = \left( \log T_{\rm eff} , V, m_1, c_1, [{\rm Fe/H}]_{\rm obs}
\right),
\end{equation}
 and we enlarged the  errors in photometry  to ten times their quoted medians; hence
\begin{equation}
  \bsigma_y = (0.01, 0.05, 0.04, 0.06, 0.1) .
\end{equation}
 We maintained the same prior as for our previous analysis, and took a flat
$\phi(\mathbf{x})$. The resultant histogram of normalized parallax residuals
is displayed in Fig.~\ref{fig:basti}, along with the mean and dispersion of
the distribution. It can be seen that although the distribution is, as ever,
unbiased, the errors in this case have been noticeably under-quoted,
resulting in a dispersion of $1.7$ in the distribution of normalized
residuals. This is hardly surprising given the questionable nature of the fit
between the models and the observations. Fig.~\ref{fig:basti_errors} shows
that the distribution of the stars in error space is similar to that of
Fig.~\ref{fig:GCS_stiletto}; however, since the top of the shoe occurs at
$\sim\! 1.4$ rather than $\sim\! 2.5$, the distance errors are now smaller,
reflecting the fact that the Str\"omgren colours give a significantly better
handle on the dwarf/giant dichotomy than standard Johnson colours.

\section{Relation to previous work} \label{sec:previouswork}

It is worth briefly considering the difference between the technique
presented here and that employed on the second RAVE data release by
\cite{Breddels}.  Breddels et al.\ estimate the distance to each star by
finding the model star that gives the smallest chi-squared value when its
observables are compared with the observations. Then 5\,000 `observed'
realizations of the model star are produced by scattering the observables of
the best-fitting star with errors drawn from a Gaussian in
$\mathbf{y}$-space, of width determined by the stated observational errors.
Finally the model that best fits each of these 5\,000 pseudo-stars is found,
and the the mean and variance of the absolute magnitudes of the 5\,001
best-fitting models is calculated.

This technique can be viewed as a form of maximum-likelihood estimation: it
essentially involves considering only the likelihood term $p(\bar{\mathbf{y}}
| \mathbf{x}, \bsigma_y )$ from our equation~(\ref{eq:final}).  The main
shortcoming of the procedure is that it fails to take advantage of the fact
that some types of star are extremely rare, so it is much more likely that a
particular datum reflects bad luck with the observations than detection of a
very rare type of star, such as a young, very metal-poor star, whose true
observable coincides with the datum.  Incorporation of the selection function
and the prior allows the computer to choose the model star that is the sanest
choice.

There is an additional problem with the procedure used by
\citeauthor{Breddels}: in the event that the data lie say $2\sigma$ from the
true values of the star's observables, the method involves fitting model
stars to data points that lie 3 or even $4\sigma$ from the true value.
Compounding errors in this way is not sensible.

\cite{Zwitter2010} have recently made significant improvements on the work
of \citeauthor{Breddels}, introducing a mass prior and a more sophisticated
treatment of the RAVE errors, as well as exploring the effects of different
sets of stellar models.

\cite{Jorgensen}, building on the work of \cite{PontEyer} (see
appendix~\ref{sec:appendix}), developed a Bayesian method for the estimation
of stellar ages (which extends naturally to other parameters), involving a
very simple prior and marginalization over all parameters except the one of
interest. The concept of regarding distance as a stellar parameter as we do
in this paper was not explored.

\citeauthor{Jorgensen} use the mode to characterize each distribution, a
decision with which we disagree for the reasons given in
Section~\ref{sec:theory}. Confidence intervals, which they used to
characterize errors, are good measures but in general comparatively difficult
to calculate, and more importantly require interpolation in the isochrones,
which we have endeavoured to avoid.

The prior used in the analysis of \citeauthor{Jorgensen} is extremely simple:
it is flat in both metallicity and age, although in initial mass it is a
single power law, representative of a simple IMF. While the rationale for
this choice is well-established, we feel that it does not take advantage of
the large body of prior knowledge that we have regarding
the samples with which we are dealing: the Galaxy is certainly not, for
example, uniformly distributed in age. It seems
logical to exploit all the information that is available, and that includes
our significant previous knowledge of the distributions of each stellar
parameter throughout the Galaxy.

An important point relevant both to the work of \citeauthor{Jorgensen} and
that of \citeauthor{Breddels} regards speed: \citeauthor{Jorgensen}'s
technique of analysing the marginalized pdf for the stellar age 
\citep[or mass in][]{Nordstrom} does not permit the versatility of calculation that our
method provides: by considering the distance as an intrinsic parameter of
each star, on the same footing as its metallicity, age and mass, we are able
to provide a consistent and simultaneous (and therefore rapid) determination
of the values of all four parameters for each star, without requiring
successive marginalizations that would significantly slow the analysis.
Likewise the technique of \citeauthor{Breddels} requires the solution of
5\,001 optimization problems for every star, as opposed to a single-pass
integration. Our technique therefore represents a significant improvement in
efficiency, which is particularly relevant in the analysis of the large
surveys that are both under way and planned.

\section{Conclusions} \label{sec:conclusions}

We have presented a Bayesian technique for determining stellar parameters
from photometric and spectroscopic data. It has been demonstrated that the
technique outperforms maximum-likelihood techniques, and the mathematical
and physical basis of the system ensures that all available information can
be exploited in the calculation. The resultant uncertainties, assuming a
reasonable prior is chosen, are therefore a consequence purely of the
underlying physics, and by virtue of this the technique is optimal -- 
given a set of data and some level of understanding
of the underlying physics, one cannot do better.  Since the
uncertainties derived from our technique simply reflect the physics,
so long as one employs a reasonable prior, the technique will provide the
most accurate possible estimates of the true uncertainties -- smaller
estimated uncertainties could only come from an overly restrictive prior,
undersampling the pdf in some manner or defining the uncertainties in a 
different manner, as for example in \cite{Jorgensen}.

The problem of fitting stellar models to observations is an ideal setting for
the application of Bayesian probability theory. We have a wealth of advance
knowledge of the underlying physics with which to construct a prior, and the
complex relationship between input parameters and observables for stellar
models renders a simple maximum-likelihood approach suboptimal for providing
reliable parameter determinations. There \textit{are} favoured regions of
observable space; our approach exploits this by virtue of the input prior.
The Bayesian method also enables a valuable synthesis of different areas of
astrophysical research: large-scale analysis and mapping of stellar
populations in the Galaxy feeds us information for the prior, whilst the
specifications of observations give us the form of the likelihood and
selection functions.

As with all photometric techniques, obscuration, which we have neglected, is
an important issue. It would be simple to include reddening and extinction
models, but inevitably the results would then be vulnerable to weaknesses in
the adopted models. When a large body of trigonometric parallaxes for
relatively distant stars is at hand, it will be possible to adjust such
models by making photometric distances compatible with trigonometric parallaxes. Such
work will be a key project to be undertaken with the Gaia Catalogue.

A simplification we have made is to consider all metallicity information to
be encoded in the value of $\mh$ for each star. The chemistry of stars is in
reality more complex and both helium abundance and alpha enhancement play
significant roles in stellar evolution. We have not explicitly addressed this issue
because to do so we would require both further sets of isochrones and
additional spectroscopic observables. However, such refinements will be
possible within a decade; most importantly, their incorporation would
represent no fundamental change to the method presented here.

While our method is strictly only applicable to single stars, dividing the
GCS sample into probable binaries and single stars reveals that the
parallaxes of binaries are under-estimated by at most $\sim\! 0.2\sigma$ in the
mean, and probably by only half this figure (Section~\ref{sec:binary}). The
effect of binarity on the estimation of other stellar parameters is more
subtle, however \cite{Jorgensen} provide a demonstration that its effect on
Bayesian age determination is quite limited.  The effect of binaries could be
incorporated into our formalism by including a probability of binarity in the
prior, along with its associated effect on the various observables; this is
the approach taken in the work of \cite{PontEyer}. Taking full account of
such contamination would require observational simulations to find the effect
of different mass-ratio binaries on each observable.

In this work we have introduced the selection function $\selfn$ as an
explicit term in our calculations.  Its inclusion provides an intuitive and
logically consistent means of taking account of all selection effects
introduced by instrumental apparatus, observing strategies and later choices
(such as cuts on errors to `clean' a sample). The presence of this factor in
our calculations in the form of $\phi(\mathbf{x})$ is key, as it truly brings
the model to the data rather than vice versa, and thus permits a
theoretically much more satisfactory basis for all our calculations.

The distribution defined by the product of the selection function and the prior
has another use. This distribution can be used as in
equation~(\ref{eq:fake}) to generate a sample of pseudodata, which can then
be scattered by Gaussian `observational errors'. The distributions of the
various observables of this pseudodata can then be compared with those of the
true data set that one wishes to mimic, in order to assess the accuracy of
the different factors one has included. Thus, if one is confident of the
selection function (as will often be the case), one can adapt the prior until a
good fit is found between the pseudodata and the real data. Once this has
been achieved, this optimized prior can then be used in the analysis of the
true data with a high degree of confidence.

In this work the form of the prior has been kept intentionally simple: we
have dealt with a case in which the Galaxy is described by three stellar
populations. Depending on the nature of the sample being analysed, the prior
could take a large range of forms. An interesting case that we have not
considered is when the data include kinematic quantities. Then the prior
would include a full phase-space distribution function for each stellar
population, and this enrichment of the prior information that is brought to
bear on the data would further reduce the uncertainties in estimated
distances, metallicities, etc. Indeed the power of the Bayesian method
described in this paper lies in its generality; although it was applied only
to spectrophotometric data in Sections~\ref{sec:pseudodata} and
\ref{sec:GCS}, nothing restricts the inclusion of other observables.

Of course the correct form for the prior is unlikely to be known absolutely.
One may however be able to specify it with a reasonable degree of certainty
-- one might, for example, be able to give the prior in a parametric form
with a certain pdf over the parameters that specify it (such as disc scale
lengths). The integration described in Section~\ref{sec:theory} can then also
be performed over these parameters, providing slightly less
assumption-dependent values for the method's outputs.

We have not explored the question of the joint probability distributions of
different parameters conditional upon the data for each star. Using the same
technique as we describe but marginalizing only over selected dimensions of
$\mathbf{x}$, it would be possible to examine such distributions and analyse
covariances between different parameters for selected stars. This extension
would be simple and one can envisage cases in which it could prove fruitful.

Another consideration regarding the pdf for each star is that of bimodality.
We have characterized each star's pdf in distance by its mean and dispersion;
however an extension of this approach would be to search through each star's
pdf in distance to identify any instances of multimodality.  Some stars
essentially present two distinct solutions for the given observables: one
being a dwarf lying comparatively nearby, the other a more distant giant (as
in the cases explored for the GCS in Section~\ref{sec:GCS}). It could be
fruitful to identify such degeneracies explicitly, and provide a mean and
dispersion for each peak in the pdf.  Starting from the algorithm described
in this paper, one could identify any points in the assigned grid whose
probabilities (by equation~\ref{eq:final}) lie above some chosen threshold
value, and then seek to identify sizeable islands in this subset of the space
by a friends-of-friends type algorithm (\citealt{Huchra}). This would provide
a method of partitioning the pdf into discrete distributions, each of which
could then be assigned an average value and associated spread. Thus
significant degeneracies in the pdf resulting from certain values of the
observables could be identified and handled suitably. We hope to explore this
avenue in a future paper.

We have shown that by using all of the available information one can constrain the metallicities of stars to greater accuracy than the
observations themselves. It is reasonable to expect that this same effect can
be achieved for other observable quantities such as the surface gravity, for
which the nominal observational errors are often sizeable. In this manner one
can use the small errors on certain observables (such as apparent magnitudes)
to shrink conservative errors on less well-constrained values. So long as one has
confidence in the stellar models used, this promises to be a powerful
technique for survey analysis. 
It should be noted that this technique can only realistically be applied once -- `looping', by rerunning the analysis using the new values and errors, is prohibited -- since its effectiveness comes from the application of independent prior knowledge to data with mildly overquoted errors. The mathematical formalism does not permit the subsequent application of a prior that is not independent of the data.

One more point deserves mention. If one knows the distances to the stars in
some specific sample (of a generic type that can be described by
model isochrones), then one can in theory explore different forms for the
prior in order to optimize one's estimation of their distances. This could
then be fed back in to the algorithm in order to provide a more accurate
estimation of distances to other stars.

We are currently using this method to obtain distances to in excess of
$250\,000$ RAVE stars and hope shortly to report the results of this work.

\section*{Acknowledgments}

Thanks are due to John Magorrian and the other members of the Oxford dynamics
group for many helpful comments on this work. BB also thanks Michael Aumer
for his initial help with isochrone handling, and acknowledges the support of
PPARC/STFC.

\bibliographystyle{mn2e}
\bibliography{refs}

\bsp

\appendix
\section[]{Jacobian questions} \label{sec:appendix}

There is some confusion in the literature regarding the necessity or
otherwise of a Jacobian term in the transition from considering the
likelihood term as a function of $\mathbf{x}$ to considering it implicitly as
a function of $\mathbf{y}(\mathbf{x})$ (such as when it is converted to the
form $G(\bar{\mathbf{y}} - \mathbf{y}(\mathbf{x}), \bsigma_y)$). The
Borel-Kolmogorov paradox (see, for example, \citealt{Jaynes}) cautions that
this step should be taken with care. However, it is
a simple matter to demonstrate that the term is unnecessary.

If we consider, for a moment, $\mathbf{y}$ not to be a function of $\mathbf{x}$ but rather to be described by a pdf on $\mathbf{x}$, we can expand equation~(\ref{eq:final}):
\begin{equation}
  p(\mathbf{x} | \bar{\mathbf{y}}, \bsigma_y , S) \propto  P(S | \bar{\mathbf{y}}, \mathbf{x}, \bsigma_y) \, p(\bar{\mathbf{y}} | \mathbf{x}, \bsigma_y ) \, p(\mathbf{x}) 
\end{equation}
\begin{equation}
   \; =  P(S | \bar{\mathbf{y}}, \mathbf{x}, \bsigma_y) 
    \left( \int \! p(\bar{\mathbf{y}} | \mathbf{y}', \mathbf{x}, \bsigma_y )
   \, p(\mathbf{y}' | \mathbf{x}) \, \dd\mathbf{y}' \right)  p(\mathbf{x}) ;
\end{equation}
but then we can express the true functional dependence of $\mathbf{y}'$  by
\begin{equation}
  p(\mathbf{y}' | \mathbf{x}) = \delta\!\left( \mathbf{y}' -
  \mathbf{y}(\mathbf{x}) \right),
\end{equation}
 leading directly to
\begin{equation}
  p(\mathbf{x} | \bar{\mathbf{y}}, \bsigma_y , S) \propto P(S | \bar{\mathbf{y}}, \mathbf{x}, \bsigma_y) \, p(\bar{\mathbf{y}} | \mathbf{y}(\mathbf{x}), \bsigma_y ) \, p(\mathbf{x})   ,
\end{equation}
which contains no Jacobian.

\label{lastpage}

\end{document}